\documentclass[trackchanges, twocolumn]{aastex701}

\usepackage{amsmath}
\usepackage{fontawesome5}
\usepackage{hyperref}
\usepackage[dvipsnames]{xcolor}
\usepackage{subcaption}

\begin{document}

\title{
    Preparing for the \textit{Gaia} DR4 Epochalypse: \\
    Simulated Epoch Astrometry for Substellar Companion Orbit-Modeling Benchmarks
}
\shorttitle{Preparing for the \textit{Gaia} DR4 Epochalypse}
\shortauthors{Yahalomi et al.}

\correspondingauthor{Daniel A. Yahalomi}
\email{yahalomi@mit.edu}

\author[0000-0003-4755-584X]{Daniel A. Yahalomi}
\altaffiliation{Juan Carlos Torres Postdoctoral Fellow}
\affiliation{Kavli Institute for Astrophysics and Space Research, Massachusetts Institute of Technology, Cambridge, MA 02139, USA}
\email{yahalomi@mit.edu}

\author[0000-0003-0872-7098]{Adrian M. Price-Whelan}
\affiliation{Center for Computational Astrophysics, Flatiron Institute, 162 Fifth Ave, New York, NY 10010, USA}
\email{aprice-whelan@flatironinstitute.org}

\author[0000-0003-3430-3889]{Jessica Speedie}
\affiliation{Department of Earth, Atmospheric, and Planetary Sciences, Massachusetts Institute of Technology, Cambridge, MA 02139, USA}
\email{jspeedie@mit.edu}

\author[0000-0001-5684-4593]{William Thompson}
\affiliation{NRC Herzberg Astronomy and Astrophysics, 5071 West Saanich Road, Victoria, BC, V9E 2E7, Canada}
\email{William.Thompson@nrc-cnrc.gc.ca}

\author[0000-0001-9907-7742]{Megan Bedell}
\affiliation{Center for Computational Astrophysics, Flatiron Institute, 162 Fifth Ave, New York, NY 10010, USA}
\email{mbedell@flatironinstitute.org}

\author[0000-0001-8183-459X]{Victor Ramirez Delgado}
\affiliation{Center for Computational Astrophysics, Flatiron Institute, 162 Fifth Ave, New York, NY 10010, USA}
\email{vramirezdelgado@flatironinstitute.org}


\author[0000-0003-0012-9093]{Aida Behmard}
\affiliation{Center for Computational Astrophysics, Flatiron Institute, 162 Fifth Ave, New York, NY 10010, USA}
\email{abehmard@flatironinstitute.org}

\author[0000-0002-9319-3838]{Avraham Binnenfeld}
\affiliation{Institute of Physics, École Polytechnique Fédérale de Lausanne (EPFL), Observatoire de Sauverny, Chemin Pegasi 51b, 1290 Versoix, Switzerland}
\email{avrahambinn@gmail.com}

\author[0000-0001-9582-4261]{Dori Blakely}
\affiliation{Department of Physics and Astronomy, University of Victoria, 3800 Finnerty Road, Elliott Building, Victoria, BC, V8P 5C2, Canada}
\affiliation{NRC Herzberg Astronomy and Astrophysics, 5071 West Saanich Road, Victoria, BC, V9E 2E7, Canada}
\email{blakelyd@uvic.ca}

\author[0000-0002-7419-9679]{Anthony G.A. Brown}
\affiliation{Leiden Observatory, Leiden University, Einsteinweg 55, 2333 CC, Leiden, The Netherlands}
\email{brown@strw.leidenuniv.nl}

\author[0000-0003-0174-0564]{Andrew R. Casey}
\affiliation{Center for Computational Astrophysics, Flatiron Institute, 162 Fifth Ave, New York, NY 10010, USA}
\affiliation{School of Physics and Astronomy, Monash University VIC 3800, Australia}
\email{acasey@flatironinstitute.org}

\author[0000-0002-3610-6953]{Jiayin Dong}
\affiliation{Department of Astronomy, University of Illinois at Urbana-Champaign, Urbana, IL 61801, USA}
\affiliation{Center for Astrophysical Surveys, National Center for Supercomputing Applications, Urbana, IL 61801, USA}
\email{jdongx@illinois.edu}

\author[0000-0001-5173-2947]{Clarissa R. Do \'O}
\affiliation{Department of Astronomy, California Institute of Technology, 1200 East California Boulevard, Pasadena, CA 91125, USA}
\email{cdoo@caltech.edu}

\author[0000-0001-9985-0643]{Caleb Lammers}
\affiliation{Department of Astrophysical Sciences, Princeton University, 4 Ivy Lane, Princeton, NJ 08544, USA}
\email{caleb.lammers@princeton.edu}

\author[0000-0003-1247-9349]{Cheyanne Shariat}
\affiliation{Department of Astronomy, California Institute of Technology, 1200 East California Boulevard, Pasadena, CA 91125, USA}
\email{cshariat@caltech.edu}

\author[0000-0002-4265-047X]{Joshua N. Winn}
\affiliation{Department of Astrophysical Sciences, Princeton University, 4 Ivy Lane, Princeton, NJ 08544, USA}
\email{jnwinn@princeton.edu}

\begin{abstract}
\textit{Gaia}'s fourth data release (DR4) will provide epoch astrometry for billions of sources, enabling the detection of potentially tens of thousands of substellar companions. We present the \texttt{epochalypse} catalog: \textit{Gaia} DR4-like epoch astrometry for testing orbit-modeling pipelines on a common set of injected systems with known parameters. Onto each of the $5{,}040{,}328$ main-sequence \textit{Gaia} stars within $500\,{\rm pc}$ in the \texttt{G23H} catalog, we inject a zero-, one-, and two-companions system, for $15{,}120{,}984$ systems in total. Companion masses and semimajor axes are drawn log-uniformly over $1\,M_{\rm Mars}$--$80\,M_{\rm Jup}$ and $0.03$--$100\,{\rm au}$, with no occurrence model and no detectability cut, so orbital periods extend well beyond the mission baseline and the companion-free population serves as a matched null. We also flag sub-populations of one- and two-companions systems whose companions are potentially detectable, based on an estimated signal-to-noise ratio that accounts for each star's noise, number of transits, and the suppression of signals with periods longer than the baseline. As a demonstration, we run the \texttt{kepmodel} astrometric periodogram on every system and find that signal detection and period localization decouple beyond the DR4 baseline: long-period companions can produce significant astrometric curvature while their periods, and therefore their masses, remain poorly constrained. Under any companion occurrence model, this population of systems where the orbit is detectable but not characterizable will far outnumber the population with fully characterizable orbits. The \texttt{epochalypse} catalog, its full generation pipeline, and a modeling tutorial are publicly available.
\href{https://doi.org/10.5281/zenodo.23789041}{\textcolor{RoyalBlue}{\faDatabase}} \href{https://github.com/dyahalomi/epochalypse}{\textcolor{RoyalBlue}{\faGithub}}
\href{https://github.com/dyahalomi/epochalypse/blob/main/tutorials/fit_astrometric_orbits.ipynb}{\textcolor{RoyalBlue}{\faGraduationCap}}
\end{abstract}



\section{Introduction \& Motivation}

\textit{Gaia} Data Release 4 (DR4), expected 2 December 2026, will for the first time include astrometric time series (epoch astrometry) sensitive to a wide range of luminous and non-luminous companions. This dataset is anticipated to deliver the largest sample of homogeneously well-characterized exoplanets to date. Astrometric companion detection yields have long been forecast to be substantial: \citet{Perryman2014} predicted $\sim$21,000 detectable giant planets over the nominal mission \citep[see also][]{Casertano2008, Sozzetti2014, Sanderson2022} and \citet{Holl2022_bd_forecast} predict $28{,}000-42{,}000$ brown dwarf detections within several hundred parsecs. Recent forecasts with an updated \textit{Gaia} noise model, modern planetary occurrence rates, and an updated stellar catalog predict $\sim$1,900 precise planetary characterizations (better than 20\% masses and periods) and $\sim$7,500 planet detections in \textit{Gaia} DR4 \citep{LammersWinn_GaiaForecast}.

In preparation for DR4, the \textit{Gaia} team has released several sample datasets: (i) per-CCD time-series astrometry for Gaia\,BH3 at the DR4 baseline fit with a preliminary version of the DR4 astrometry fitting pipeline \citep{GaiaBH3_2024}; (ii) the ``OHP'' sample dataset,\footnote{\url{https://dace.unige.ch/openData/?record=10.82180/dace-gaia-ohp}} seven injected benchmark systems spanning diverse scenarios (a single star, a scan-angle artifact, high- and low-SNR orbits near $\sim$90 and $\sim$$2{,}000$\,d, and a long-period acceleration solution) with DR4-like per-epoch astrometry; and (iii) the DR4 prerelease,\footnote{\url{https://www.cosmos.esa.int/web/gaia/dr4-prerelease}} providing \textit{actual} DR4 per-CCD measurement epoch astrometry for 12 systems, three of which have non-single-star orbital solutions (Gaia\,BH3, Gaia-4, and HD\,114762). 

In anticipation, the community has developed a range of astrometric orbit-modeling packages and modeling methods -- including \texttt{orbitize!} \citep{Blunt2020_orbitize, Blunt2024_orbitize}, \texttt{orvara} \citep{Brandt2021_orvara}, \texttt{Octofitter} \citep{Thompson2023_octofitter}, \texttt{htof} \citep{Brandt2021_htof}, \texttt{kima} \citep{Faria2018_kima, Baycroft2026_kima}, the \texttt{exoplanet}/\texttt{jaxoplanet} ecosystem \citep{ForemanMackey2021_exoplanet_JOSS, ForemanMackey2021_exoplanet_zenodo, jaxoplanet}, and \texttt{harv}\footnote{\url{https://github.com/adrn/harv}} \citep[Price-Whelan, in prep.; built on the same rejection-sampling approach as \texttt{The\,Joker},][]{PriceWhelan2017_joker}. These packages span different likelihood formulations, prior distributions, parameterizations, and sampling strategies. There exists no catalog of injected systems spanning a wide range of masses, periods, and multiplicities, which would enable a comparison of these approaches directly, and separate what comes from the data from what comes from the sampling method and modeling choices.

We address these needs with the \texttt{epochalypse} catalog: injected planetary and substellar (non-luminous) systems with realistic, \textit{Gaia} DR4-like epoch astrometry. We exclude luminous stellar companions where the photocenter motion of near-equal-mass unresolved binaries is expected to be the dominant astrophysical false positive in astrometric exoplanet detection \citep{Holl2023_DR3orbits, MarcussenAlbrecht2023, Stefansson2025, Barbato2026,
Marcussen2026}, but such systems are likely identifiable with photometric, machine-learning, or radial velocity vetting \citep[see e.g.,][]{Wallace2024, Chance2025, SahlmannGomez2025, BailerJones2026}.

Injection-recovery catalogs most naturally adopt one of two approaches in selecting their underlying injection population: (1) attempt to best reproduce the intrinsic underlying population based on current understanding of the population providing forecasted yields or (2) assume no occurrence rate but rather attempt to inject a population spanning all potential areas of parameter space enabling tests of sampler precision, accuracy, and recovery completeness. Here, we adopt the latter approach as there are already several catalogs for \textit{Gaia} astrometric planets optimized on the former approach \citep[see e.g.,][]{Sozzetti2014, Perryman2014, Holl2022_bd_forecast, Sanderson2022, LammersWinn_GaiaForecast}, but create a large enough catalog of systems such that users can downselect based on an assumed occurrence rate. The \texttt{epochalypse} catalog contains a zero-, one-, and two-companions system injected onto every main-sequence star in the \texttt{G23H} catalog out to 500 pc. The \texttt{epochalypse} catalog contains a population of substellar companions that exhaustively spans all physical regimes of possible parameter space, reflected by the following principles:

\begin{enumerate}
    \item \textbf{Agnostic to the parent population.} We sample from broad distributions in companion mass, semimajor axis, and eccentricity, spanning the full range of non-luminous, substellar companion masses from terrestrial bodies ($1\,M_{\rm Mars}$) to the hydrogen-burning limit ($80\,M_{\rm Jup}$). This range deliberately extends below the expected detection threshold. We do not attempt to predict detailed stability criteria for two-planet interactions, but rather inject spanning wide ranges and follow the analytic two-planet stability criteria \citep{Gladman1993}, additionally retaining the more tightly packed pairs that lie within $5\%$ of a $2{:}1$ or $3{:}2$ mean-motion resonance, where resonant protection is possible. Users can adopt additional stability cuts as they see fit.

    \item \textbf{Zero-companions, one-companion, and two-companions.} Spanning multiplicities -- including companion-free noise realizations -- enables tests of both statistical false-positive rates and the ability to disentangle multiple companions. Here, statistical false positives are those induced by sampler bias or features inherent to the \textit{Gaia} scanning law \citep[see e.g.,][]{Holl2023_scanlaw}, as opposed to the astrophysical false positives discussed above.

    \item \textbf{Cross-comparison of methods.} We release the injected time series themselves, enabling direct, end-to-end pipeline comparison. Because the injected parameters are known exactly, this permits an absolute assessment of accuracy, establishing where modeling choices agree and where biases induced by sampling choices may arise.
\end{enumerate}

Together, these design choices make the \texttt{epochalypse} catalog a flexible foundation for orbit-modeling, completeness and statistical false-positive studies, and the systematic benchmarking of \textit{Gaia} astrometric planet-recovery methods. All the code used to generate the \texttt{epochalypse} catalog is modular and made public, such that its injection framework could also be used to inject substellar companions onto real \textit{Gaia} DR4 epoch astrometry and noise, once the data is released. The description of the catalog creation is divided in three main parts: Section~\ref{sec:stars} describes the process of building the parent stellar population used in the catalog and the noise model; Section~\ref{sec:companions} describes the process of injecting the companion populations onto the parent stellar population; Section~\ref{sec:astrometry} describes the astrometric model adopted to generate the \textit{Gaia} DR4-like per-epoch astrometry in the \texttt{epochalypse} catalog. Section~\ref{sec:results} presents a simple periodogram based recovery analysis of all \texttt{epochalypse} systems, compares the \texttt{epochalypse} recovered systems to those in the \citet{LammersWinn_GaiaForecast} catalog, and discusses the differences between the recovered companions with ``characterized orbits'' and those with only ``detected orbits''; and finally, Section~\ref{sec:use} presents the data structure of the \texttt{epochalypse} catalog and suggests a range of potential use cases.

\section{Parent Stellar Population} \label{sec:stars}

\begin{figure*}
\centering
\includegraphics[width=\textwidth]{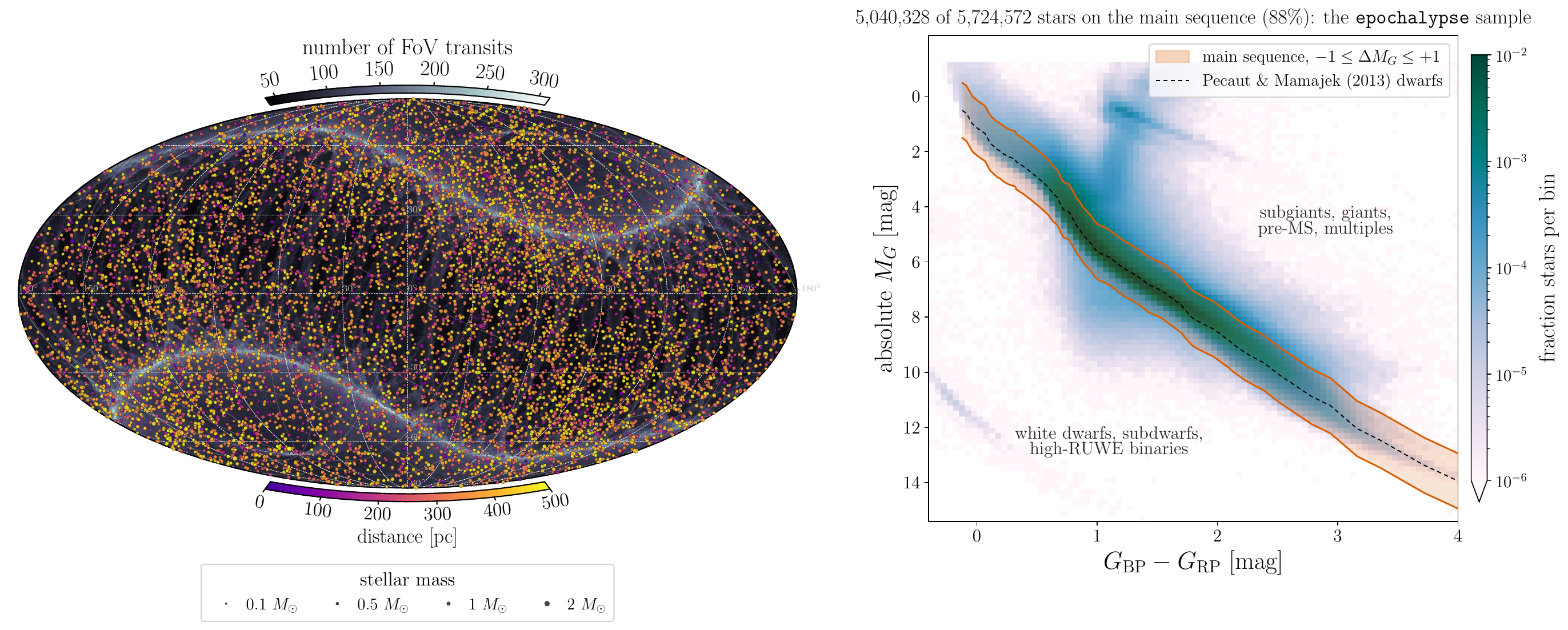}
\caption{\emph{Left:} $10{,}000$-star random subset over the predicted DR4 per-epoch Gaia scanning law map, colored by distance and scaled in marker area by stellar mass. Note this is the number of FoV transits without the $\sim$10\% expected data loss. \emph{Right:} two-dimensional histogram showing the fraction of \texttt{epochalypse} stars in each bin of the color--magnitude diagram ($G_{\rm BP}-G_{\rm RP}$ against $M_G$). The dashed line is the \citet{PecautMamajek2013} mean dwarf sequence. The mass and radius of every star in \texttt{epochalypse} are interpolated from this sequence using its absolute magnitude $M_G$. The shaded band, $-1 \le \Delta M_G \le +1$ about the \citet{PecautMamajek2013} mean dwarf sequence, marks the stars we identify as main sequence stars (MS): $5{,}040{,}328$ of $5{,}724{,}572$. Magnitudes and colors are not corrected for extinction, which could move reddened MS dwarfs above the band.}
\label{fig:star_sky_distance}
\end{figure*}

\subsection{Extracting Stars from \texttt{G23H} Catalog}
The parent stellar population underlying the \texttt{epochalypse} catalog is constructed from the \texttt{G23H} catalog \citep{Thompson_G23H, Blakely_G23H}. The \texttt{G23H} catalog contains all stars with a \textit{Gaia} DR2--DR3 ID crossmatch. We restrict the parent stellar sample to all stars within $500\,{\rm pc}$ (parallax $\varpi \geq 2\,{\rm mas}$). As the astrometric signal of a companion is larger for nearby stars, this 500~pc cut is motivated by the combination of computational feasibility and the expectation that the majority of \textit{Gaia} DR4 astrometric companion detections will arise from nearby systems. This choice is consistent with recent yield forecasts suggesting that fewer than $10\%$ of \textit{Gaia} DR4 planet detections originate beyond $250\,{\rm pc}$ \citep{LammersWinn_GaiaForecast} -- though for the more massive substellar companions, this distance threshold should not hold.

Applying a $500\,{\rm pc}$ heliocentric distance cut to the \texttt{G23H} catalog returns $5{,}789{,}521$ entries. \texttt{G23H} carries one row per DR2--DR3 cross-match, and $60{,}419$ DR3 sources appear more than once ($60{,}392$ twice and $27$ three times). Because these duplicated rows differ only in their DR2 cross-match columns we collapse each group to a single row, retaining the cross-match with the smallest DR2 source identifier so that the choice is deterministic and independent of row ordering. This downselection results in the unique DR3 subset of $5{,}729{,}075$ stars. Additionally, we remove $1{,}094$ stars that have missing noise values that we require in the \texttt{G23H} catalog, resulting in $5{,}727{,}981$. All stars in the \texttt{G23H} parent catalog have a \textit{Gaia} $G$-band magnitude brighter than $16$, ensuring well-characterized astrometry across the sample.

\subsection{Interpolating Stellar Masses and Radii}
\label{sec:pecaut}

We estimate stellar masses and radii from the empirical dwarf sequence tabulated by \citet{PecautMamajek2013}. For each source, we calculate the absolute \textit{Gaia} $G$-band magnitude ($M_G$) using the apparent $G$ magnitude and parallax from the \texttt{G23H} values taken from \textit{Gaia} DR3. We then use the resulting $M_G$ value to interpolate the tabulated stellar mass $M_\star$ and radius $R_\star$. We performed no color-based cross-match.

We restrict interpolation to the range of $M_G$ over which the \citet{PecautMamajek2013} dwarf sequence tabulates absolute $G$ magnitude, mass, and radius simultaneously, without extrapolation. The sequence itself extends from O3V to L8V, but $M_G$ is tabulated only for spectral types B3V and later, and stellar mass only for L2V and earlier; the overlap spans $-1.19 \le M_G \le 17.30$, corresponding to $5.4\,M_\odot$ ($3.61\,R_\odot$, B3V) at the luminous end and $0.075\,M_\odot$ ($0.097\,R_\odot$, L2V) at the faint end. $3{,}409$ stars fall outside this range; excluding them leaves $5{,}724{,}572$ stars. 

We additionally remove stars that do not lie on the main sequence in the \textit{Gaia} color-magnitude diagram. For each star we compute the offset from the dwarf sequence, $\Delta M_G = M_G - M_{G,{\rm MS}}(G_{\rm BP}-G_{\rm RP})$, where $M_{G,{\rm MS}}$ is the \citet{PecautMamajek2013} absolute $G$ magnitude interpolated at the star's observed color along the B9V--M8.5V ridge, and remove stars with $\Delta M_G < -1.0$ or $\Delta M_G > +1.0$. Of the $5{,}724{,}572$ stars in the parent sample, we remove $622{,}252$ above the band (giants, subgiants, pre-main-sequence stars, reddened dwarfs, and higher-order multiples), $59{,}956$ below it (white dwarfs, subdwarfs, and high-RUWE binaries), and $2{,}036$ outside the color range of the ridge, leaving $5{,}040{,}328$ stars. The resulting stellar population has parallaxes spanning $2.00$--$546.98\,{\rm mas}$, a median magnitude of $G = 14.04$, stellar masses of $0.078$--$4.12\,M_\odot$, and includes $44{,}307$ stars carrying \textit{Hipparcos} identifiers. Assigning each star the \citet{PecautMamajek2013} spectral type whose tabulated $M_G$ is nearest its own, the sample is dominated by K dwarfs ($2{,}769{,}064$ stars, 55\%), followed by G ($930{,}288$, 18\%), early-M (M0--M3.5; $733{,}190$, 15\%), and F dwarfs ($530{,}212$, 11\%); mid-M dwarfs (M4--M6.5) contribute $25{,}796$ stars (0.5\%), late-M dwarfs (M7 and later) only $28$, and B and A stars together $51{,}750$ (1\%). Figure~\ref{fig:star_sky_distance} shows the color--magnitude diagram of the full parent stellar catalog, binned in $G_{\rm BP}-G_{\rm RP}$ and absolute \textit{Gaia} $M_G$ and colored by the fraction of the sample falling in each bin. Bins are $0.05$~mag in color by $0.2$~mag in $M_G$ magnitude.

\subsection{Obtaining \textit{Gaia} Scanning Law}\label{sec:scanlaw1}
We obtained the \textit{Gaia} scanning law via queries of the \textit{Gaia} Observation Forecast Tool (\texttt{GOST}).\footnote{\url{https://gaia.esac.esa.int/gost/index.jsp}} For each epoch of a given sky position, \texttt{GOST} returns the barycentric observation time $t_{\rm TCB}$, the along-scan position angle $\psi$, the along-scan parallax factor $\Pi$, and the field of view (preceding or following). Querying \texttt{GOST} involves a relatively slow web query, so following \citet{El-Badry2024}, we first download the \textit{Gaia} scanning law (observational epochs plus additional measurement information) at a given location in the sky, by querying \texttt{GOST} at the center of each pixel of a level 6 HEALPix grid in equatorial coordinates comprising 49,152 pixels of $0.916^{\circ}$ resolution. This is comparable to \textit{Gaia}'s $0.7^{\circ}$ across-scan FoV. We query the full DR4 baseline, spanning 2014 July 25 10:30 UTC to 2020 January 20 22:00 UTC (2005 days, or 66 months of nominal mission operations). 

Due to a variety of issues \citep[see][]{Lindegren2021_GaiaEDR3_astrometric}, about 10\% of FoV transits do not result in usable data \citep{El-Badry2024, LammersWinn_GaiaForecast}. The DR4 prerelease epoch astrometry offers an additional test of this prediction (see Section~\ref{sec:noise_test}). The 12 stars in the prerelease dataset have a median value of 11\% (range 4--33\%) of epochs with no CCD observations used by the Astrometric Global Iterative Solution (AGIS; \citealt{Lindegren2012_AGIS}) -- further supporting this $\sim$10\% unusable epochs prediction. We thus randomly reject 10\% of all epochs per-star to reflect this expected data loss. This yields a median of $81$ epochs per star (range $32$--$293$).\footnote{Each star inherits its pixel's scanning law and loses a random 10\% of its epochs; this ignores sub-pixel variations in scanning law and the clustering of real data gaps in time, but is an appropriate approximation for population-level statistics.}

Figure~\ref{fig:star_sky_distance} shows the equatorial coordinates and distances of a random subsample of $10{,}000$ stars in the stellar catalog, overlaid on a Gaia scanning law map of \textit{Gaia} field-of-view (FoV) transit counts over the DR4 baseline for reference.


\subsection{Noise model} \label{sec:noise}

\textit{Gaia} astrometric observations will be reported at the CCD level (up to 9 `astrometric field' CCD measurements per epoch), each with its own measurement uncertainty. We anticipate that (i) the community will be motivated -- primarily by computational efficiency -- to combine the per-CCD measurements into a single per-epoch measurement prior to orbit modeling, and that (ii) a deeper understanding of the noise properties, and of how to combine the CCD measurements gracefully, will come in time with active research. Here, we aim to devise a conservative noise model that embodies current wisdom pre-DR4 and reflects the anticipated drive to work with per-epoch measurements.

Following the findings by \citet{Kiefer2025}, we report analogous per-epoch uncertainties as were used in DR3, but simulate the uncertainties inferred by \citet{Kiefer2025}, which include systematics not captured in the \textit{Gaia} team's noise model used for DR3. Our noise model thus has two parts: the scatter we add to the simulated astrometry, and the uncertainty we report alongside each measurement. The two are built from different combinations of the \citet{Kiefer2025} noise terms. The per-CCD uncertainty is split into three components: the along-scan measurement uncertainty $\sigma_{\rm AL}$, the spacecraft attitude excess noise $\sigma_{\rm att}$, and the calibration noise $\sigma_{\rm cal}$. We take all three from the \texttt{G23H} catalog \citep{Thompson_G23H}, where they were computed following \citet{Kiefer2025} for both five- and six-parameter solutions. We also use the mean number of along-scan CCD windows per transit,
\begin{equation}
\langle N_{\rm AL}\rangle = \frac{N_{\rm AL}}{N_{\rm FoV}} .
\end{equation}

In short, the formal uncertainties reported by \textit{Gaia} underestimate the true scatter: single stars in DR2 and DR3 have a unit weight error (UWE) above one, which motivated the empirical renormalization to the renormalized unit weight error (RUWE) of \citet{Lindegren2018_RUWE_dr2}. We therefore adopt the ``unbiased estimator of variance a posteriori'' (UEVA) of \citet[][Sect.~5.1.1]{Kiefer2025}: the variance of the along-scan residuals of the five-parameter astrometric fit, per CCD measurement. ${\rm UEVA}_{\rm single}$ is the unbiased estimator of variance a posteriori for a single star. That work showed ${\rm UEVA}_{\rm single}$ reproduces the single star residual scatter measured by \citet{Lindegren2021_GaiaEDR3_astrometric}, so it is the noise expected for a single star on top of which any companion-induced excess appears. The difference between the measured ${\rm UEVA}_{\rm ruwe}$ (or ${\rm UEVA}_{\rm aen}$) and ${\rm UEVA}_{\rm single}$ is their indicator of non-singleness. Following Equation 25 of \citet{Kiefer2025}, we compute the expectation value of ${\rm UEVA}_{\rm single}$ for each star, in units of mas$^{2}$ per CCD measurement, as
\begin{equation}\label{eq:mu_UEVA_single}
\begin{split}
\mu_{\rm UEVA,single} &= \frac{\langle N_{\rm AL}\rangle}{N_{\rm AL} - n_{\rm dof}} \\
&\quad \times \left[(N_{\rm FoV} - n_{\rm dof})\,\sigma_{\rm cal}^2
      + N_{\rm FoV}\,\sigma_{\rm AL}^2\right],
\end{split}
\end{equation}
following the \texttt{G23H} implementation of \citet{Thompson_G23H}, with $n_{\rm dof}$ the number of parameters fitted in the DR3 astrometric solution, which is removed from the residual degrees of freedom:
\begin{equation}
n_{\rm dof} =
\begin{cases}
5, & \texttt{astrometric\_params\_solved\_dr3}=31,\\
6, & \text{otherwise.}
\end{cases}
\end{equation}
From Equation~\ref{eq:mu_UEVA_single}, ${\rm UEVA}_{\rm single}$ depends on the calibration and AL noise. To convert this to a noise value in mas per epoch, we take the square root of $\mu_{\rm UEVA,single}$ -- the regression standard error -- as an estimate of the typical error of a per-CCD measurement \citep[following][]{Blakely_G23H}, and naively scale by $\langle N_{\rm AL}\rangle$:
\begin{equation}\label{eq:sigma_UEVA_single}
\begin{split}
\sigma_{\rm UEVA,single}
&= \sqrt{\frac{\mu_{\rm UEVA,single}}{\langle N_{\rm AL}\rangle}} \\
&= \sqrt{\frac{(N_{\rm FoV} - n_{\rm dof})\,\sigma_{\rm cal}^2
             + N_{\rm FoV}\,\sigma_{\rm AL}^2}{N_{\rm AL} - n_{\rm dof}}} \, .
\end{split}
\end{equation}
This $\sigma_{\rm UEVA,single}$ is the per-epoch scatter we use to generate the simulated astrometry in \texttt{epochalypse}.

For the uncertainty reported with each measurement we use
\begin{equation}\label{eq:sigma_formal}
\sigma_{\rm formal} = \sqrt{\frac{\sigma_{\rm att}^2 + \sigma_{\rm AL}^2}{\langle N_{\rm AL}\rangle}},
\end{equation}
which combines the attitude and AL terms but leaves out the calibration term of Equation~\ref{eq:sigma_UEVA_single}, mirroring the DR3 formal errors \citep{Kiefer2025}. That is, we inject the scatter we expect a single star to intrinsically have, and report the uncertainty we expect to be reported in the DR4 catalog.

Finally, we let the measurement quality vary from epoch to epoch. In the DR4 prerelease epoch astrometry (Section~\ref{sec:noise_test}), the standard deviation of a source's per-epoch $\sigma_{\rm formal}^{\rm DR4}$ (Equation~\ref{eq:dr4_epoch_sigma}) across its epochs is 4--22\% of its median (median 7\%) for the eight sources with $G \le 16$. We adopt a 10\% fractional perturbation. For each epoch $i$ we draw $\xi_i \sim \mathcal{N}(0,1)$ and apply the same 10\% fractional perturbation to both the injected and reported uncertainties,
\begin{equation} \label{eq:sigma_injected}
\sigma_{{\rm injected},i} = (1+0.1\,\xi_i)\,\sigma_{\rm UEVA,single},
\end{equation}
\begin{equation} \label{eq:sigma_reported}
\sigma_{{\rm reported},i} = (1+0.1\,\xi_i)\,\sigma_{\rm formal},
\end{equation}
and generate the astrometric residual as
\begin{equation}
\epsilon_i \sim \mathcal{N}\!\left(0,\sigma_{{\rm injected},i}^2\right).
\end{equation}
We use the same $\xi_i$ for both so that an epoch with larger injected scatter is also assigned a proportionally larger formal uncertainty, while the ratio between the injected scatter and the uncertainty supplied to the fit is preserved.

We note that bright stars ($G \lesssim 13$) carry additional systematic errors in DR3 from gated, often saturated windows \citep{Lindegren2021_GaiaEDR3_astrometric, ElBadry2021_GaiaEDR3}. This degradation is accounted for in our noise model through the magnitude dependence of the \citet{Kiefer2025} injected noise terms that increase the simulated noise around brighter stars ($G \lesssim 13$) as shown in Figure~\ref{fig:dr4_prerelease_sigma}.

\subsubsection{Testing the Noise Model} \label{sec:noise_test}

\begin{figure}
    \centering
    \includegraphics[width=\columnwidth]{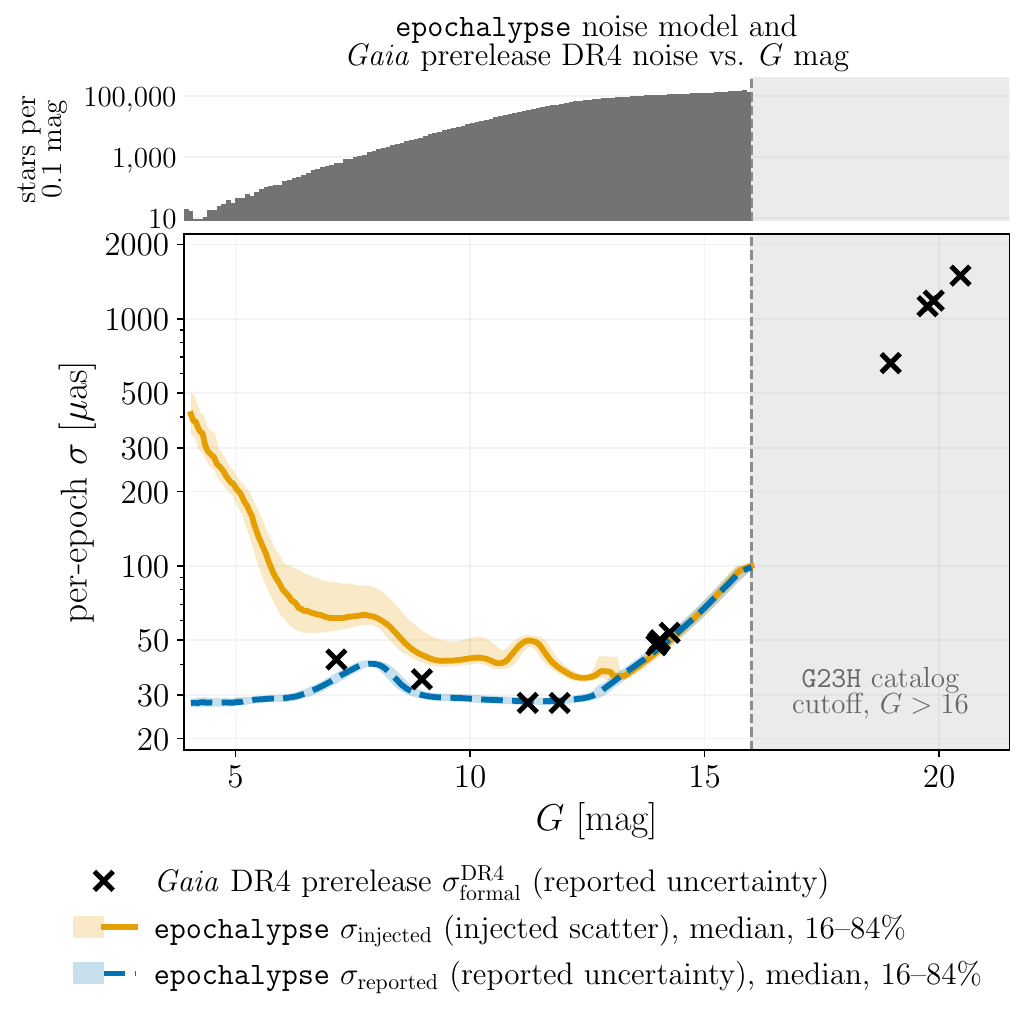}
    \caption{Per-epoch along-scan uncertainty against $G$ magnitude. Crosses mark the 12 \textit{Gaia} DR4 prerelease systems, their per-CCD errors combined per epoch through Equation~\ref{eq:dr4_epoch_sigma} and the median plotted. Lines show the moving median and $16$--$84\%$ range over the $5{,}040{,}328$ stars of the parent \texttt{epochalypse} sample of the injected scatter $\sigma_{\rm injected}$ (Equation~\ref{eq:sigma_UEVA_single}; yellow, solid) and the reported uncertainty $\sigma_{\rm reported}$ (Equation~\ref{eq:sigma_reported}; blue, dashed). The top panel is the \texttt{epochalypse} sample's $G$ distribution. Beyond $G=16$ (shaded) lies outside the \texttt{G23H} limit, so the 4 faintest systems are shown for context alone.}
    \label{fig:dr4_prerelease_sigma}
\end{figure}

The \textit{Gaia} DR4 prerelease -- \textit{actual} DR4 per-CCD measurement epoch astrometry for 12 systems -- provides a check of our assumed noise model.

Being reported per-CCD, the DR4 prerelease data first require combination to per-epoch astrometry to compare to the per-epoch noise model reported in the \texttt{epochalypse} catalog. Each epoch carries ten measurements -- the sky-mapper (SM) window, used for object identification rather than astrometry, followed by the nine astrometric-field windows AF1--AF9. We keep all measurements flagged by \texttt{used\_by\_agis\_al == True}, which retains $7{,}467$ of the $10{,}080$ ($74\%$). The discarded quarter is of two kinds: the SM measurements, never flagged and so a full tenth of the total by construction ($1{,}008$ total), and a further $1{,}605$ astrometric-field measurements that AGIS itself rejected ($16\%$ of all measurements, $18\%$ of the AF measurements). Those sharing a \texttt{transit\_id} are then combined into a single epoch by an inverse-variance weighted mean, each CCD carrying weight $w_i = 1/\sigma_i^2$ with $\sigma_i \equiv \texttt{centroid\_pos\_error\_al}$, so that the uncertainty on the combined epoch is the standard error of that mean,
\begin{equation}\label{eq:dr4_epoch_sigma}
\sigma_{\rm formal}^{\rm DR4} = \left(\sum_i \sigma_i^{-2}\right)^{-1/2} \, .
\end{equation}
If per-CCD errors were equal within a given epoch this reduces to $\sigma_i/\sqrt{\langle N_{\rm AL}\rangle}$, and is therefore the same construction as our $\sigma_{\rm formal}$. In total there are $54$--$93$ epochs per system.

8 of the 12 prerelease stars are within the \texttt{G23H} magnitude limit of $G=16$ and so for those 8 stars, $\sigma_{\rm formal}^{\rm DR4}$ can be compared against the median of the $\sigma_{\rm reported}$ we assign as a function of $G$-magnitude (Figure~\ref{fig:dr4_prerelease_sigma}). The ratio of $\sigma_{\rm formal}^{\rm DR4}$ to our median predicted $\sigma_{\rm reported}$ at a given magnitude has a median value of $1.06$ and spans $0.98$--$1.19$. The upper end of that spread is set by the two brightest stars ($G=7.2$ and $G=9.0$), which sit $16$--$19\%$ high; for the remaining six, all at $G>10$, the median ratio is $1.05$ and the span narrows to $0.98$--$1.07$. While 8 systems is a small subset and insufficient to fully validate a noise model, they suggest that the reported uncertainties from our noise model appear correct at the $\sim$10\% level.

\subsection{Summary}
The construction of the stellar parent population thus comes from combining several sources. From the \texttt{G23H} catalog \citep{Thompson_G23H} we adopt the star's \textit{Gaia} DR3 parallax $\varpi$ and proper motions $(\mu_{\alpha\star}, \mu_\delta)$ \citep{GaiaMission2016, GaiaDR3_2023}, along with the DR3 noise terms ($\sigma_{\rm AL}$, $\sigma_{\rm att}$, $\sigma_{\rm cal}$) following the noise model described in \citet{Kiefer2025} and as applied in \citet{Blakely_G23H} (Section~\ref{sec:noise}). The noise model can be seen compared to the 12 \textit{Gaia} DR4 prerelease sources in Figure~\ref{fig:dr4_prerelease_sigma}. Main sequence stars are identified and the stellar mass and radius values are interpolated from the \citet{PecautMamajek2013} main sequence stellar values (Section~\ref{sec:pecaut}). Finally, via \texttt{GOST} query at the star's HEALPix pixel we take the observational epochs $t_{\rm TCB}$, the along-scan scan angle $\psi$, the along-scan parallax factor $\Pi$, and the field-of-view identifier.

\section{Injected Companion Populations} \label{sec:companions}

\begin{figure*}
\centering
\includegraphics[width=\textwidth]{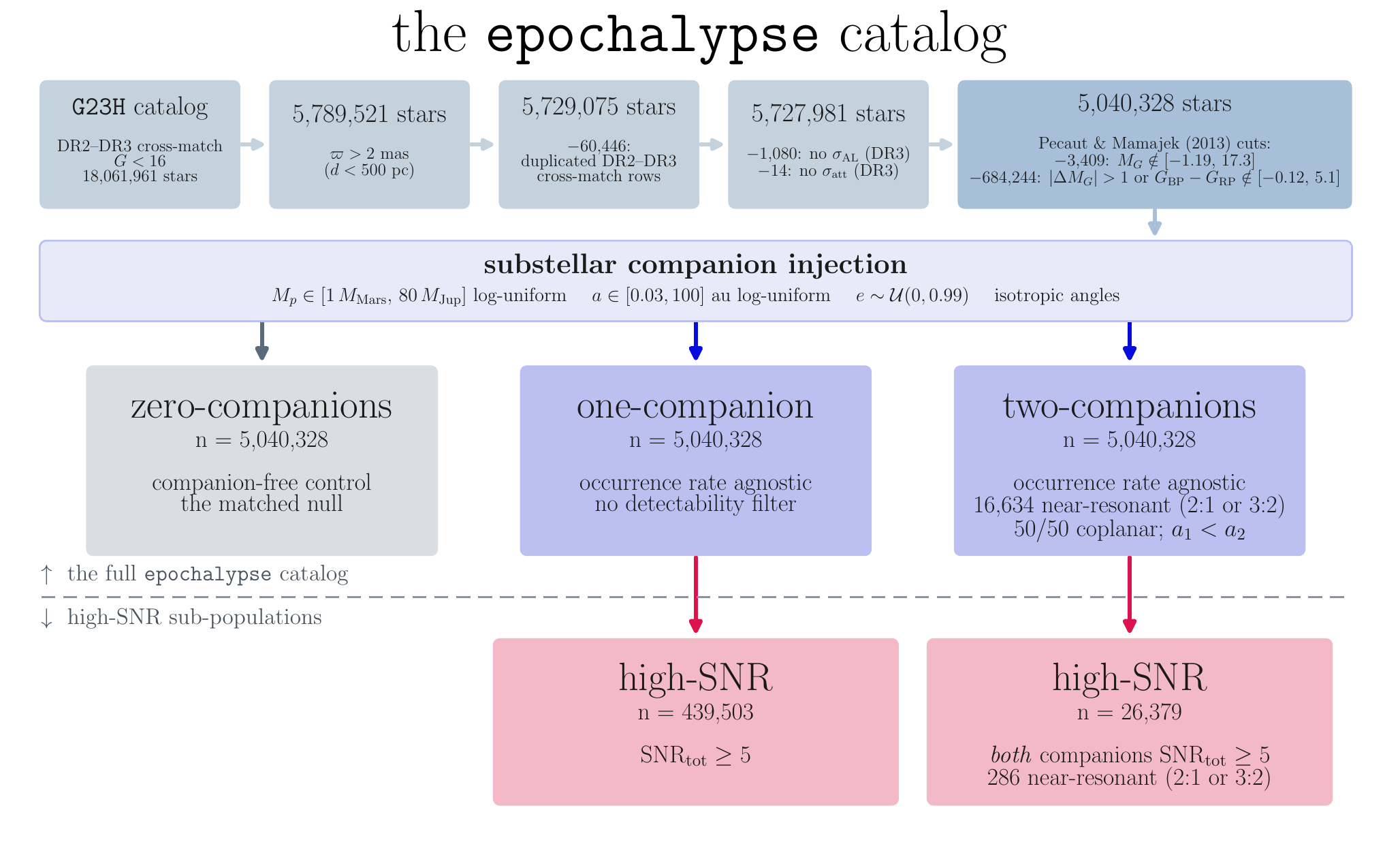}
\caption{Construction of the \texttt{epochalypse} catalog. \emph{Top:} the stellar down-selection (Section~\ref{sec:stars}) from the full \texttt{G23H} release to a parent sample of $5{,}040{,}328$ main-sequence stars within 500~pc. \emph{Middle:} onto each star, we inject three populations -- zero-, one-, and two-companions -- and simulate \textit{Gaia} DR4-like astrometry for each system. \emph{Bottom:} the two high-SNR sub-catalogs are selections over the populations above rather than separate simulations, admitting the $439{,}503$ and $26{,}379$ systems whose companions all clear $\mathrm{SNR_{tot}} \geq 5$.}
\label{fig:schematic}
\end{figure*}

We inject zero-, one-, and two-companions populations onto the parent stellar sample. The zero-companions population provides a catalog as if every star in the catalog had no companions, our one-companion population provides a catalog as if every star in the catalog had a single companion, and the two-companions population provides a catalog as if every star in the catalog had two companions. Thus, with $5{,}040{,}328$ stars in the parent stellar population, we produce $3 \times (5{,}040{,}328) = 15{,}120{,}984$ total systems in the \texttt{epochalypse} catalog. A schematic showing the five populations of the \texttt{epochalypse} catalog can be seen in Figure~\ref{fig:schematic}.

\subsection{The Agnostic Populations}\label{subsec:generated_pops}

\subsubsection{Zero-Companions}\label{subsec:zero-companion}
The ``zero-companions'' population, as the name suggests, contains no injected orbital companion onto the parent stellar population. This provides a noise-only null model with the full observational structure of the survey and none of the companion-induced signal. Any signal is by construction a false positive, so the population delivers the statistical false-alarm distribution of whatever detection statistic is adopted. Detection thresholds calibrated on this population can then be applied unchanged to the one- and two-companions populations to give completeness at a fixed statistical false-alarm rate. We note that the injected noise is white and Gaussian by construction and does not contain expected false positives such as photocenter shifts caused by unresolved stellar binaries -- so statistical false-alarm rates measured here should be read as a floor for the real DR4 epoch data rather than a prediction of it.

\subsubsection{One-Companion}\label{subsec:one-companion-random}

The ``one-companion'' population is designed to broadly span physically plausible orbital architectures. In this population, for each star in the stellar parent population, a single planetary companion is assigned by independent random draws of its mass, orbit, and orientation, with no detectability cut applied. Companion masses are drawn from a broad distribution spanning terrestrial to brown-dwarf companion masses ($1\,M_{\rm Mars} \leq M_p \leq 80\,M_{\rm Jup}$). Both the mass and the orbital semimajor axis are drawn log-uniformly, i.e., uniform in $\log_{10} M_p$ over the range above and uniform in $\log_{10} a$ over $0.03\,{\rm au} \leq a \leq 100\,{\rm au}$. Eccentricities are drawn from a uniform distribution ($e \sim \mathcal{U}(0,\,0.99)$).

Inclinations are drawn uniformly in $\cos i$ (equivalently $\cos i \sim \mathcal{U}(-1,1)$, so $i \in [0^\circ, 180^\circ]$), while the argument of periastron $\omega$ and the longitude of the ascending node $\Omega$ are each drawn uniformly over $[0, 2\pi)$. We draw the mean anomaly at the reference epoch, $M_0 \equiv M(t_0)$ with $t_0 = \mathrm{JD}\ 2457936.875$ (TCB), uniformly over $[0, 2\pi)$, and convert it to a time of periastron passage $T_p = t_0 - M_0 P / 2\pi$, which is passed to the orbit integrator.

The orbital period follows from Kepler's third law, retaining the companion mass since it is non-negligible for the highest-mass companions in the distribution (an $80\,M_{\rm Jup}$ brown dwarf is $\sim$15\% of a $0.5\,M_\odot$ star, and a larger fraction for the lowest-mass hosts):
\begin{equation}
P^2 = \frac{4\pi^2 a^3}{G\,(M_\star + M_p)}.
\end{equation}

The corresponding astrometric signature of each system is the angular reflex semimajor axis of the host,
\begin{equation}
\alpha = \left(\frac{M_p}{M_\star + M_p}\right) a\, \varpi
       = \left(\frac{M_p}{M_\star + M_p}\right)\frac{a}{d},
\end{equation}
where $a$ is the orbital semimajor axis, $\varpi$ is the parallax, and $d$ is the distance to the system. In practice $\alpha$ is computed directly from the parallax ($a\,\varpi$ with $a$ in au and $\varpi$ in mas yields $\alpha$ in mas).

The mass vs. period space of companions in the \texttt{epochalypse} catalog can be seen in Figure~\ref{fig:simulated_planets_mass_period}. The one-companion population parameter distributions are shown as blue shaded histograms in Figure \ref{fig:1planet_pop}.

\subsubsection{Two-Companions}\label{subsec:two-companions-random}

For each star in the ``two-companions'' population, two substellar companions are drawn independently, each from the identical per-companion distributions used above. We note that we treat each companion as an independent Keplerian, summed linearly, and do not model companion-companion interactions.

Rather than impose a physically motivated mutual inclination distribution, we aim to cover both limiting cases. The mutual orientation of the two companions is set by a $50/50$ mixture: with probability one-half the system is drawn coplanar and coaligned, the outer companion sharing the inner companion's inclination and node ($i_2 = i_1$, $\Omega_2 = \Omega_1$); otherwise the two inclinations and nodes are drawn independently and isotropically. We label the companions so that companion~1 is the inner ($a_1 < a_2$) and companion~2 the outer.

Because two closely spaced companions assigned in this manner are often dynamically unstable, each candidate pair is passed through a stability filter and, if rejected, re-drawn in full. A pair is rejected if the orbits cross, i.e., if the inner apoapsis reaches the outer periapsis,
\begin{equation}
a_1(1+e_1) \;\ge\; a_2(1-e_2).
\end{equation}
Non-crossing pairs are evaluated by their separation in mutual Hill radii,
\begin{equation}
\begin{split}
\Delta &= \frac{a_2 - a_1}{R_{\rm H}}, \\
R_{\rm H} &= \left(\frac{\mu_1 + \mu_2}{3}\right)^{1/3}\frac{a_1 + a_2}{2}, \\
\mu_k &= \frac{M_{p,k}}{M_\star}.
\end{split}
\label{eq:hill}
\end{equation}
Pairs with $\Delta \ge 2\sqrt{3}$ -- the analytic two-planet Hill-stability boundary \citep{Gladman1993} -- are retained as stable. 

More tightly packed pairs ($\Delta < 2\sqrt{3}$) are rejected as likely unstable unless they lie near a $2{:}1$ or $3{:}2$ mean-motion resonance (MMR), in which case they are retained as potentially resonantly protected. The choice of retaining systems near a $2{:}1$ or $3{:}2$ MMR is motivated by the fact that these are the two commensurabilities near which an excess of planet pairs is observed, both in the \textit{Kepler} period-ratio distribution, where the excess sits a few percent wide of exact resonance \citep[see e.g.,][]{Lissauer2011, Fabrycky2014, SteffenHwang2015}, and among radial-velocity giant planets, where several pairs are confirmed to be librating in resonance \citep[see e.g.,][]{Marcy2001, Correia2009, Wright2011}. Proximity to resonance is measured from the period ratio $P_2/P_1$, retaining systems within a fractional tolerance of $5\%$ of the $2{:}1$ or $3{:}2$ commensurabilities,
\begin{equation} \label{eq:mmr}
\left|\frac{P_2/P_1}{(j+1)/j} - 1\right| < 0.05,
\qquad j \in \{1, 2\}.
\end{equation}

Near-resonant systems are kept in the catalog as (1) resonant interactions can stabilize otherwise unstable orbital configurations and (2) the astrometric reflex motion of a single, moderately eccentric companion is degenerate (to first-order in eccentricity) with two companions in a $2{:}1$ resonant orbit \citep{Yahalomi_resoeccentric}: the \textit{resoeccentric degeneracy}. Thus, it is important to include near-resonant systems in a \textit{Gaia} benchmark catalog, so that users can investigate the ability of their samplers to differentiate between single eccentric and $2{:}1$ resonant companions over different regimes of parameter space. We note that these near-resonant systems are placed near commensurability but are not migrated into resonance, as was done in \citet{Greenfield2026}. These systems are thus not guaranteed to be resonantly protected or dynamically stable. Migration into resonance would be required in practice for resonance induced stability, but this approximation affords us a computationally inexpensive approximation of what those systems may look like in DR4 data.

In total, $16{,}634$ near-resonant systems are retained in the \texttt{epochalypse} catalog two-companions population, with $12{,}422$ near 2:1 and $4{,}212$ near 3:2 resonance.

We adopt these criteria as a minimal stability screen and do not claim that every system in the two-companions population is dynamically stable. The $2\sqrt{3}$ boundary is derived for initially circular, coplanar orbits \citep{Gladman1993}, whereas our companions are drawn with $e \sim \mathcal{U}(0, 0.99)$ and, in half of systems, isotropic mutual inclinations. Moreover, Hill stability is weaker than Lagrange stability as it forbids close encounters between the companions but not the slow escape of the outer one \citep{MarchalBozis1982, BarnesGreenberg2006}. More rigorous stability criteria exist: for example the AMD-based Hill criterion of \citet{Petit2018}, the empirical two-planet boundary of \citet{Petrovich2015}, the resonance-overlap chaos criterion of \citet{HaddenLithwick2018}, stability of hierarchical systems \citep{MardlingAarseth2001}, and stability in mutually inclined systems \citep{BhaskarPerets2024}. Because all orbital elements are tabulated, any of these or other stability criteria can be applied as a post-hoc cut to the \texttt{epochalypse} catalog, which we leave to the user in keeping with the principle of being agnostic to the parent companion population.

\begin{figure*}
\centering
\includegraphics[width=.9\textwidth]{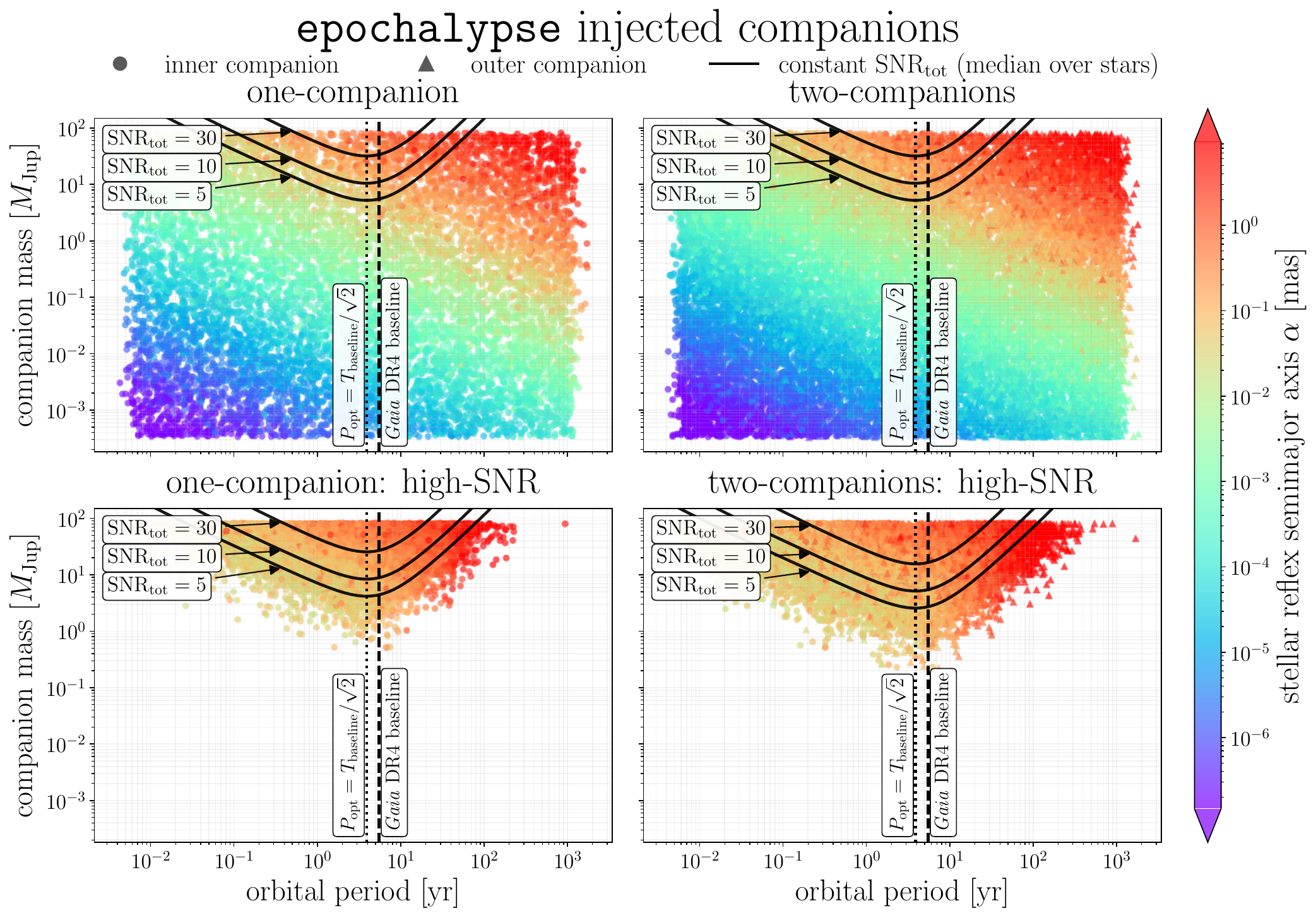}
\caption{Masses, periods, and sky projected reflex semimajor axes for the [top]: one-companion \& two-companions populations and [bottom]: their respective high-SNR sub-populations in the \texttt{epochalypse} catalog (sub-sample of $10{,}000$ systems plotted per population). The right columns show both companions, with the inner companion a circle and the outer companion a triangle. Contours show the median location of the $\mathrm{SNR}_{\rm tot}$ for the stars in the \texttt{epochalypse} catalog. Stars extend well below the $\mathrm{SNR}_{\rm tot}=5$ contour in the high-SNR population because the contours are drawn for the median star; thus half the stars, with lower noise, more transits, or larger parallax, reach $\mathrm{SNR}_{\rm tot}=5$ at lower companion mass. Vertical lines show the \textit{Gaia} DR4 baseline (5.5~yr) and $P_{\rm opt} = T_{\rm baseline}/\sqrt{2} \simeq 3.9\ {\rm yr}$.}
\label{fig:simulated_planets_mass_period}
\end{figure*}

\begin{figure*}
\centering

\begin{subfigure}[t]{0.49\textwidth}
    \centering
    \includegraphics[width=\linewidth]{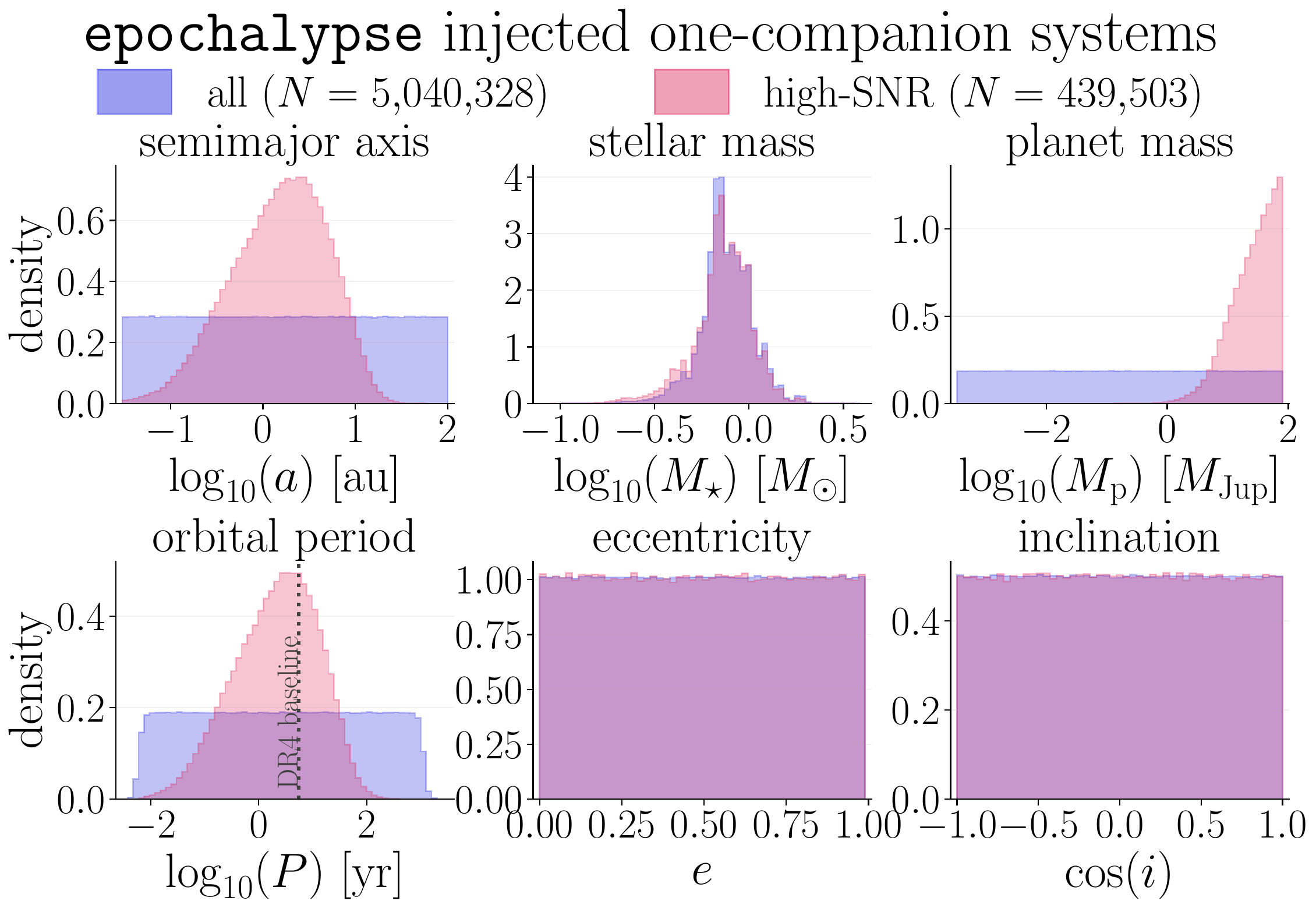}
    \caption{Orbital parameters of one-companion population.}
    \label{fig:1planet_pop}
\end{subfigure}
\hfill
\begin{subfigure}[t]{0.49\textwidth}
    \centering
    \includegraphics[width=\linewidth]{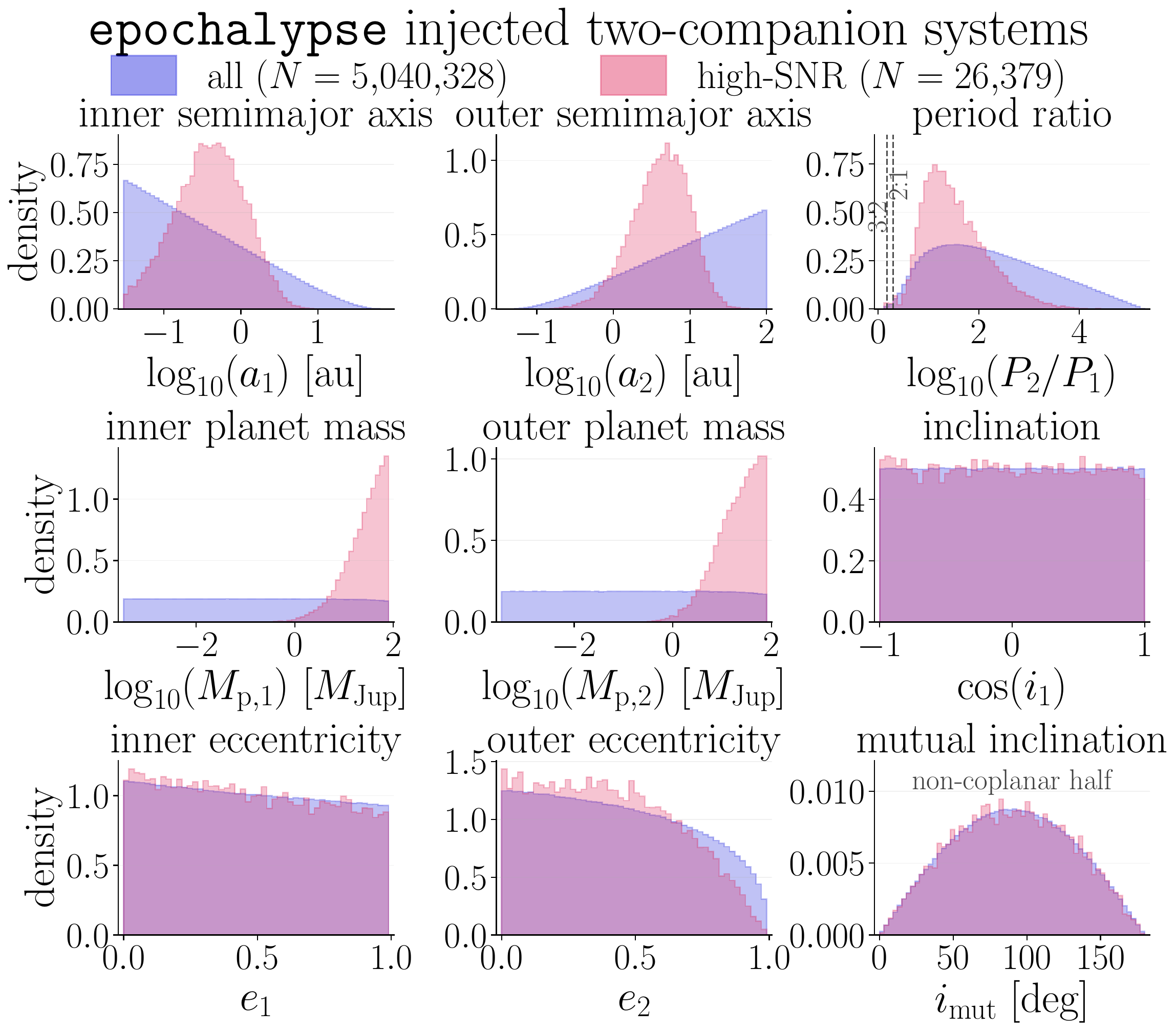}
    \caption{Orbital parameters of two-companions population.}
    \label{fig:2planet_pop}
\end{subfigure}

\caption{Orbital parameters of the one-companion (left) and two-companions (right) populations in the \texttt{epochalypse} catalog. Also the orbital parameters of the high-SNR sub-populations for both the one-companion (left) and two-companions (right).}
\label{fig:planet_pops}
\end{figure*}

The mass vs. period space of companions in the \texttt{epochalypse} catalog can be seen in Figure~\ref{fig:simulated_planets_mass_period}. The two-companions population parameter distributions are shown as blue shaded histograms in Figure \ref{fig:2planet_pop}.

\subsection{The High-SNR Sub-Populations}
For users who are interested primarily in companions expected to produce detectable signals, we also label a sub-population of the agnostic one-companion and two-companions populations as the high-SNR population. In order to do so, we first discuss expected signal-to-noise ratio (SNR) of \textit{Gaia} astrometric data.

\subsubsection{SNR of Exoplanet Astrometry}
Following \citet{LammersWinn_GaiaForecast}, we begin from SNR of a single epoch,
\begin{equation}
{\rm SNR}_1 = \frac{\alpha}{\sigma_{\rm UEVA, single}},
\qquad
\alpha = \frac{M_p}{M_\star + M_p}\, a\, \varpi ,
\label{eq:snr1}
\end{equation}

\noindent
where $\alpha$ is the maximum amplitude of the star's angular reflex astrometric displacement, $\varpi$ is the parallax, and $\sigma_{\rm UEVA, single}$ is the per-epoch injected uncertainty.\footnote{The injected epoch uncertainties carry an additional 10\% per-epoch jitter about this value, and the catalog reports the smaller formal uncertainty $\sigma_{\rm formal}$ (\S\ref{sec:noise}).}

A large reflex amplitude does not guarantee that a planet is detectable, specifically in the case where the companion's orbital period exceeds the observing baseline \citep[see e.g.,][]{Eisner2001astrometry, Gould2008}. In this case, the host star traverses only a portion of its reflex ellipse. If the host star's observed arc can be well represented as an offset in position and proper motion, then the five-parameter astrometric solution can absorb the companion signal, the same finite-baseline limit that sets the Rayleigh resolution of a periodogram \citep[see e.g.,][]{RamirezDelgado2025}. Thus, it is the curvature of the arc -- or its acceleration -- that distinguishes a detectable signal in the long period regime. \citet{Gould2008} showed that the sensitivity to astrometric companions falls off rapidly once the period exceeds $\sim$0.7 times the baseline. We can account for this by suppressing the signal-to-noise ratio of long-period orbits with an effective detectability metric that we will show scales with acceleration at long periods,
\begin{equation}
\begin{split}
{\rm SNR}_{\rm eff}
&= \frac{{\rm SNR}_1}{1 + \left(P/T_{\rm baseline}\right)^2}
 = \frac{{\rm SNR}_1}{1 + \left(a/a_{\rm crit}\right)^3}, \\
a_{\rm crit} &= \left(T_{\rm baseline}^2\, (M_\star+M_p)\right)^{1/3},
\end{split}
\label{eq:snreff}
\end{equation}

\noindent
where $T_{\rm baseline} = 5.5\,{\rm yr}$ is the \textit{Gaia} DR4 time baseline and $a_{\rm crit}$ is the semimajor axis whose orbital period equals $T_{\rm baseline}$. The two forms are equivalent by Kepler's third law.

For $P \ll T_{\rm baseline}$ the denominator approaches unity and ${\rm SNR}_{\rm eff} \to {\rm SNR}_1$, so short period orbits are not penalized. For $P \gg T_{\rm baseline}$, ${\rm SNR}_{\rm eff} \simeq {\rm SNR}_1\,(T_{\rm baseline}/P)^2$. The power of two here comes from the displacement of an accelerating source growing as $t^2$ \citep{Brandt2024}, so that the curvature of the arc is a fraction $\sim (T_{\rm baseline}/P)^2$ of the full amplitude. Since ${\rm SNR}_1 \propto \alpha \propto (M_p/(M_\star+M_p))\, a\, \varpi$ (Eq.~\ref{eq:snr1}), in this $a \gg a_{\rm crit}$ limit, ${\rm SNR}_{\rm eff} \propto (M_p\,\varpi/a^2)\,T_{\rm baseline}^2$, the acceleration of the star due to the companion times the square of the time over which it acts.

Detectability depends on the full astrometric time series. The number of epochs a star is observed is set by the \textit{Gaia} scanning law and varies strongly across the sky, so we include this cadence in our star-by-star selection. For a source observed $N$ times, assuming independent epoch measurements, the cumulative signal-to-noise grows as $\sqrt{N}$, giving a total SNR
\begin{equation}
{\rm SNR}_{\rm tot} = \sqrt{N}\;{\rm SNR}_{\rm eff}
= \frac{\sqrt{N}\,{\rm SNR}_1}{1 + \left(P/T_{\rm baseline}\right)^2}.
\label{eq:snrtot}
\end{equation}
We take $N$ directly from the predicted per-star \textit{Gaia} DR4 scanning law, i.e.\ the number of FoV transits forecast for each individual target (median $N \approx 81$ including the $\sim$10\% loss). We note that $N$ grows roughly linearly with mission duration, so Equations~\ref{eq:snreff} and \ref{eq:snrtot} together give the $T_{\rm baseline}^{5/2}$ scaling of acceleration sensitivity with baseline described in \citet{Brandt2024}.

The solid curves in Figure~\ref{fig:simulated_planets_mass_period} are contours of constant ${\rm SNR}_{\rm tot}$. We evaluate Equation~\ref{eq:snrtot} for each star and plot the median, i.e., the companion mass and orbital period at which half the sample reaches a given signal, rather than substituting median stellar parameters into the formula. At short periods, the companion mass to period relationship reduces to $M_p \propto P^{-2/3}$. The period suppression in Equation~\ref{eq:snreff} acts in the opposite direction and dominates at long periods, and there the companion mass to period relationship rises as $M_p \propto P^{4/3}$. A contour of constant signal is therefore a V-shape in the mass--period plane rather than a single power law. The two effects balance at
\begin{equation}
P_{\rm opt} = T_{\rm baseline}/\sqrt{2} \simeq 3.9\ {\rm yr},
\label{eq:popt}
\end{equation}

\noindent
so the most favorable orbits for \textit{Gaia} have periods a little inside the observing baseline. We did not tune Equation~\ref{eq:snreff} to produce this value and it is thus encouraging that the optimal period in our final SNR equation $P_{\rm opt}/T_{\rm baseline} = 1/\sqrt{2} \simeq 0.71$ approximately matches the $\sim$0.7 ratio at which \citet{Gould2008}, in an independent Fisher-matrix analysis, found astrometric parameter errors to begin growing faster than the naive $\sqrt{N}$ scaling.

\subsubsection{One-Companion: High-SNR}\label{subsec:one-companion-high-SNR}

We classify a companion as high-SNR if ${\rm SNR}_{\rm tot} \ge 5$ -- this is a somewhat arbitrary threshold that we take to mean ``is potentially detectable'' and corresponds to the often-used 5-$\sigma$ detection threshold. We note that if users want to adopt a different threshold, they are free to do so from the standard agnostic populations. As a reference, \citet{LammersWinn_GaiaForecast} found that in their catalog of injected planetary signals, systems in \textit{Gaia} DR4 with ${\rm SNR}_1 \gtrsim 1.5$ were roughly detectable. They did not apply a long period penalty, but only injected planets out to 7~yrs. For short period systems and the median star in \texttt{epochalypse} with $\sim$81 observations, the \citet{LammersWinn_GaiaForecast} threshold is equivalent to an ${\rm SNR}_{\rm tot} \gtrsim 13.5$.

The one-companion high-SNR sub-population is comprised of $439{,}503$ systems that met this ${\rm SNR}_{\rm tot} \ge 5$ threshold. The mass vs. period space of companions in the \texttt{epochalypse} catalog can be seen in Figure~\ref{fig:simulated_planets_mass_period}. Their parameter distributions are shown as red shaded histograms in Figure \ref{fig:1planet_pop}.

\subsubsection{Two-Companions: High-SNR}\label{subsec:two-companions-high-SNR}

The two-companions: high-SNR population applies the same ${\rm SNR}_{\rm tot} \ge 5$ sub-population criterion to both planets in a given system in the ``two-companions'' population. The two-companions high-SNR sub-population is comprised of $26{,}379$ systems. In total, 286 near-resonant systems are retained in the two planet high-SNR sub-population, with 215 near 2:1 and 71 near 3:2 resonance. The mass vs. period space of companions in the \texttt{epochalypse} catalog can be seen in Figure~\ref{fig:simulated_planets_mass_period}. Their distributions are shown as red shaded histograms in Figure \ref{fig:2planet_pop}.

\section{\textit{Gaia} DR4-like Simulated Astrometry} \label{sec:astrometry}

We generate \textit{Gaia}-like per-epoch astrometry for every star--companion system in Section~\ref{sec:stars} and Section~\ref{sec:companions}.

\subsection{Astrometric model}

The noiseless along-scan abscissa at each epoch is the sum of a standard five-parameter astrometric term and the companion-induced reflex term. Following the local-plane-coordinate convention of \citet{LindegrenBastian2022}, the five-parameter contribution is
\begin{equation} \label{eq:5param}
w_{\rm 5par} =
\left(\Delta\alpha^\star + \mu_{\alpha\star}\,t\right)\sin\psi
+ \left(\Delta\delta + \mu_\delta\,t\right)\cos\psi
+ \varpi\,\Pi ,
\end{equation}
where the reference-position offsets $(\Delta\alpha^\star, \Delta\delta)$ are set to zero (positions referenced to the catalog value) and all terms are in mas.\footnote{See the local-plane-coordinate description from the \textit{Gaia} technical note GAIA-C3-TN-LU-LL-061 \citep{LindegrenBastian2022}.}

The reflex motion is computed from a full Keplerian orbit model with \texttt{jaxoplanet} \citep{jaxoplanet}. Each companion is treated as an independent two-body problem sharing the same central star, and the stellar reflex is the superposition of the individual contributions; mutual companion--companion interactions are neglected.

\texttt{jaxoplanet} evaluates positions in a Cartesian frame centered on the system barycenter whose $(x, y)$ plane is the tangent plane of the sky at the star, and whose $z$-axis points from the barycenter \emph{toward} the observer, following the convention of the transit and radial-velocity literature. In this frame $(x, y)$ is the displacement of the star from the system barycenter, evaluated as the sum over all bodies,

\begin{equation} \label{eq:reflex_xy}
(x, y) \;=\; -\sum_{k=1}^{N_{\rm c}} \frac{M_{p,k}}{M_\star + M_{p,k}}\,\left(x, y\right)_{{\rm rel},k} ,
\end{equation}

\noindent
where $N_{\rm c}$ is the number of companions and $\left(x, y\right)_{{\rm rel},k}$ is the sky-plane position of companion $k$ relative to the star, evaluated at that epoch from its Keplerian elements $(P_k, e_k, i_k, \omega_k, \Omega_k, M_{0,k})$: each companion pulls the star opposite its own displacement, with the mass ratio setting the amplitude. The displacement, returned in units of $R_\odot$, is converted to an angular reflex amplitude through the parallax,

\begin{equation}
(x, y)_{\rm mas} = (x, y)_{R_\odot}\,\frac{R_\odot}{\mathrm{au}}\,\varpi ,
\end{equation}

\noindent
where $R_\odot/\mathrm{au}$ is the solar radius expressed in astronomical units, evaluated numerically from the constants in \texttt{astropy} \citep{astropy:2013, astropy:2018, astropy:2022}, $R_\odot/\mathrm{au} = 0.00465047$.

Classical astrometry -- the Thiele-Innes and Campbell element sets adopted for Equation~(\ref{eq:5param}) and by \textit{Gaia} -- defines the ascending node as the node at which the companion \emph{recedes} from the observer, that is, with the line-of-sight axis reversed relative to \texttt{jaxoplanet}. A given numerical $(\Omega, \omega)$ therefore describes orbits rotated by $180^\circ$ on the sky in the two conventions. Reading the injected elements in the classical sense, the north and east offsets of the stellar reflex are

\begin{equation}
\left(\Delta\delta,\ \Delta\alpha^{\star}\right) = -\left(x, y\right)_{\rm mas} , \qquad \Delta\alpha^{\star} \equiv \Delta\alpha\cos\delta ,
\end{equation}

and projecting onto the along-scan direction in the same sense as the five-parameter term of Equation~(\ref{eq:5param}), $w = \Delta\alpha^{\star}\sin\psi + \Delta\delta\cos\psi$, gives

\begin{equation} \label{eq:reflex}
w_{\rm reflex} = -\left(x_{\rm mas}\cos\psi + y_{\rm mas}\sin\psi\right) .
\end{equation}

The overall sign is thus a conversion between node conventions, and the injected $(\Omega, \omega)$ are on the standard astrometric convention throughout. The total model abscissa is $w_{\rm true} = w_{\rm 5par} + w_{\rm reflex}$, and the simulated measurement is $w_{\rm obs} = w_{\rm true} + \varepsilon$ with $\varepsilon \sim \mathcal{N}(0, \sigma_{\rm injected,i}^2)$ (see Section~\ref{sec:noise}).

\subsection{Reproducibility and data products}

All randomness is seeded deterministically: a master seed is combined with each system's identifier through a \texttt{blake2s} hash to yield a stable per-system seed, which is split into independent streams for the uncertainty realization and the observation noise, so any system can be regenerated exactly.

For each simulated system we write a \textit{Gaia} DR4 epoch-astrometry (OHP-like) table with one row per epoch $t_{\rm TCB}$, the along-scan centroid position $w_{\rm obs}$, its reported formal uncertainty $\sigma_{\rm reported}$, the parallax factor $\Pi$, the scan angle $\psi$, the field-of-view identifier, and the \textit{Gaia} DR3/DR2 source identifiers. In parallel we record an injected-solutions table listing, for every system, the true injected parameters (parallax, proper motions, stellar mass and radius, and each companion's $a$, $e$, $i$, $\Omega$, $\omega$, $M_0$, and $M_p$) together with all seeds, enabling exact recovery tests. The per-system epoch tables and the injected-solutions table are further explained in Section~\ref{sec:use_data}.

\section{Results} \label{sec:results}

The epoch astrometry of a given DR4 source can constrain a companion in one of three qualitatively different ways. We will refer to a companion as \textit{orbit characterized} if the astrometry both establishes its presence and localizes its period and mass. We will refer to it as \textit{orbit detected} if the signal is significant but the orbital characteristics are not recovered, so that the system is unlikely to be a single star while its period and mass remain weakly constrained and potentially multimodal. And we will refer to it as \textit{noise-dominated} if no significant signal is present, a regime that calibrates the statistical false-positive rates and sets the upper limits that any completeness correction requires. 

We emphasize that the \texttt{epochalypse} catalog contains no planetary or brown-dwarf occurrence model. Companion masses and separations are drawn log-uniformly from a deliberately broad distribution (Section~\ref{sec:companions}), so the injected populations are not samples of the Galaxy. Translating such a catalog into a yield requires an occurrence prescription of the kind adopted by \citet{LammersWinn_GaiaForecast}, which we do not attempt here, but which could be done with \texttt{epochalypse} natively via importance sampling.

\subsection{Periodogram characterization of \texttt{epochalypse}}
\label{sec:results_periodogram}

\begin{figure*}
\centering
\includegraphics[width=\textwidth]{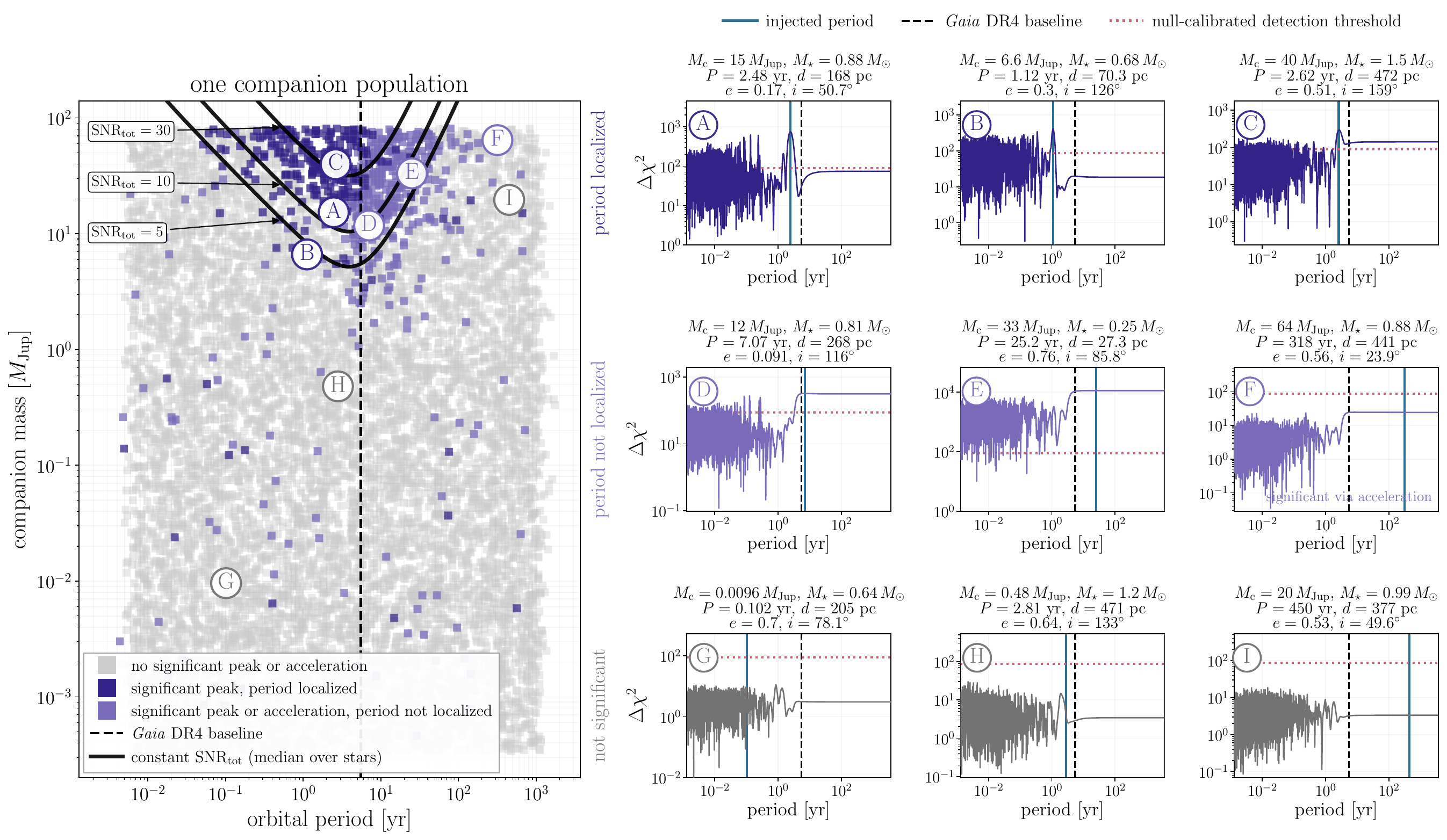}
\caption{Nine systems drawn from the one-companion population, lettered on the map at left and shown as periodograms at right, three per class. The dotted red line is the detection threshold, $\Delta\chi^2 = 88.4$, set so that 0.5\% of the 5{,}040{,}328 companion-free control systems exceed it (the other half of the 1\% false-positive budget goes to the acceleration test). Localized periods (top row) occur predominantly inside the DR4 baseline; beyond it the $\Delta\chi^2$ rises into a plateau at the long-period end of the grid.}
\label{fig:examples}
\end{figure*}

\begin{figure*}
\centering
\includegraphics[width=\textwidth]{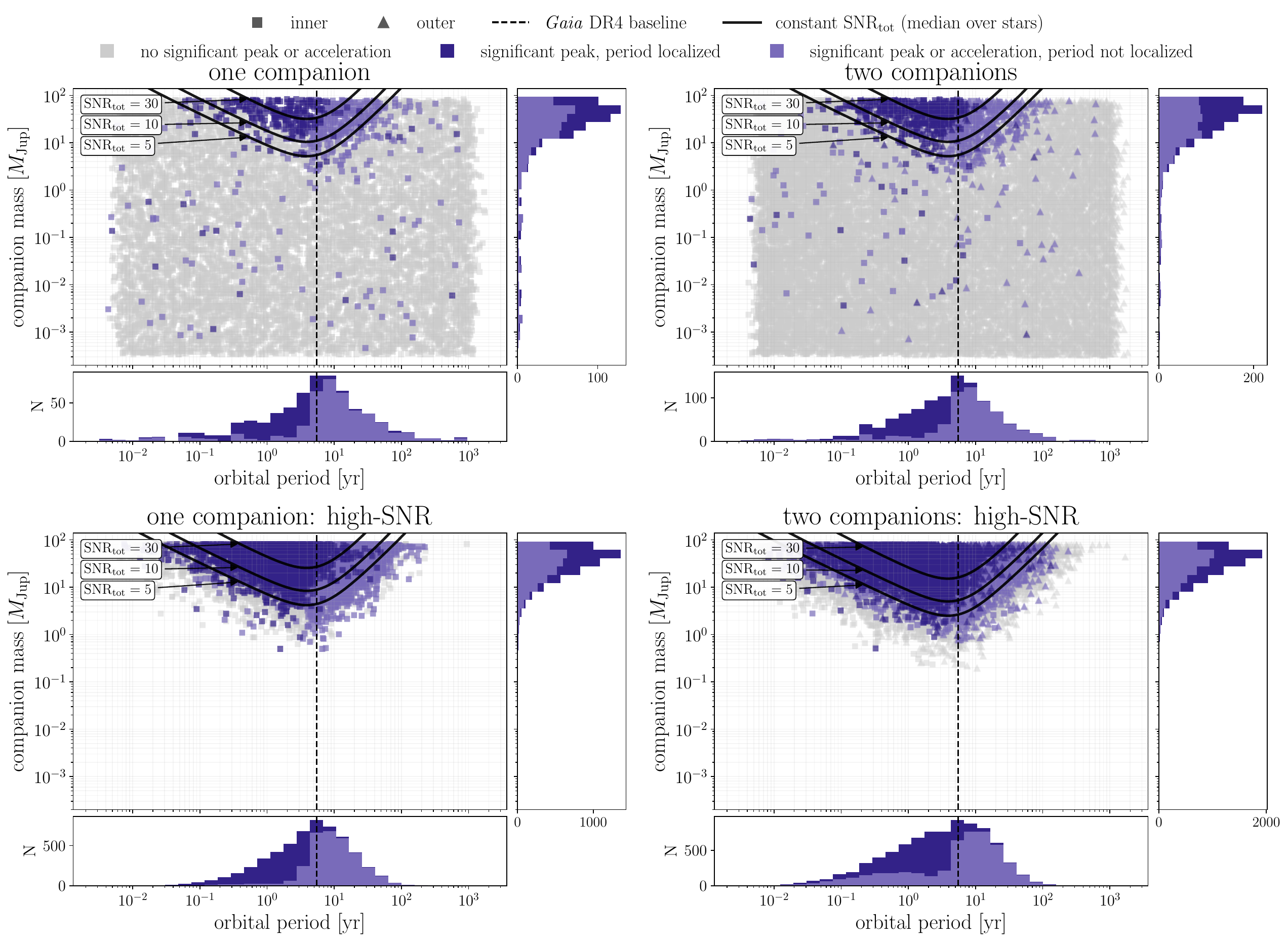}
\caption{Mass--period plane showing recovery of the four injected populations. Each companion is colored by what the \texttt{kepmodel} periodogram establishes about its period, using thresholds calibrated on the companion-free control at a $1\%$ false-positive rate. Solid curves show constant $\mathrm{SNR}_{\rm tot}$ values, evaluated per star and shown at the median value.}
\label{fig:characterizability}
\end{figure*}

Period search in \textit{Gaia} astrometry is a distinct problem from conventional time-series analysis. Since each measurement is a one-dimensional projection along a changing scan direction, the scan geometry determines both how the true astrometric signal is encoded in the data and how scan-angle-dependent noise and systematics can appear, making careful treatment of the scan geometry essential. The periodic signal must also be disentangled from position, proper motion, and parallax, which can absorb part of the orbital motion. The irregular sampling and changing scanning angles of \textit{Gaia} can further produce source-dependent signals and, in some cases, spurious periods linked to the scanning law \citep{Holl2023_scanlaw}. 
 
Several approaches have been developed for this problem. \citet{Delisle2022_kepmodel} introduced \texttt{kepmodel}, a linear astrometric periodogram that fits the standard astrometric parameters together with a periodic orbital signal. \citet{Binnenfeld2023} proposed a model-independent alternative based on phase-distance correlation, which does not assume a specific form for the periodic signal. These methods make different trade-offs between physical assumptions, sensitivity, and computational cost. Here we use the \texttt{kepmodel} periodogram as a straightforward and efficient diagnostic.

We fit every system using the astrometric periodogram implemented in \texttt{kepmodel}\footnote{\url{https://gitlab.unige.ch/jean-baptiste.delisle/kepmodel}} \citep[the periodogram used in the \textit{Gaia}/OHP tutorial;][]{Delisle2022_kepmodel}. In short, at each trial frequency the model fits the five linear astrometric columns of Eq.~(\ref{eq:5param}) together with four Thiele-Innes columns for a circular ($e=0$) orbit, and reports the improvement $\Delta\chi^2 = \chi^2_{\rm 5par} - \chi^2(P)$ relative to the companion-free model. Because the model is linear in all nine coefficients, they can be profiled out analytically at each trial period -- i.e., set to their best-fitting values, which weighted linear least squares gives in closed form. $\chi^2(P)$ is then the lowest $\chi^2$ attainable at that period, and the search is a profile-likelihood scan over the single non-linear parameter, $P$. 

The choices we make are as follows. (i) The measurement weights are held fixed at $1/\sigma_{\rm reported}^2$, with no excess-noise term fitted, so that the statistic tests the astrometry as it is reported. (ii) The trial grid comprises $15{,}777$ periods spanning $1.3\times10^{-3}$--$3{,}700\,{\rm yr}$, built as $32$ log-period segments each searched on its own uniform frequency grid, giving $\Delta\log_{10}P = 3.3$--$5.2\times10^{-4}$ throughout. (iii) A second, independent channel tests for acceleration alone, as the $\Delta\chi^2$ of a seven-parameter model (five-parameter plus along-scan acceleration) against the five-parameter one. (iv) Both channels are thresholded empirically on the companion-free control population, splitting a $1\%$ total false-positive budget evenly between them, which gives $\Delta\chi^2_{\rm orbit} > 88.4$ and $\Delta\chi^2_{\rm accel} > 18.9$; $0.89\%$ of zero-companions systems clear either channel. Calibrating on the companion-free control -- the same $5{,}040{,}328$ stars as the one- and two-companions populations -- lets us set these thresholds empirically, from a noise-only sample, rather than assuming an analytic null distribution.

We call a period localized when two conditions hold. (i) The competitive region must be narrow: taking the set of trial periods whose $\Delta\chi^2$ falls within $4$ of the maximum, we require its total extent in $\log_{10}P$ to be less than $0.05$~dex, a factor of $1.12$ in period, corresponding to a period determined to about $\pm 6\%$ when the region is contiguous. (ii) The peak must also be isolated and interior to the period grid limitations: no second distinct peak, separated by at least $0.1$~dex, may lie within $\Delta\chi^2 = 10$ of the tallest, and the maximum may not sit within $2\%$ of the log-period range of either end of the grid.

Every companion is then assigned to one of three classes, which are the classes plotted in Figures~\ref{fig:examples} and \ref{fig:characterizability}:

\begin{itemize}
  \item \textbf{No significant peak or acceleration.} Neither channel clears its 1\% false positive threshold. The astrometry is consistent with a single star, and the companion sits in the noise-dominated regime.

  \item \textbf{Significant peak or acceleration, period not localized.} At least one channel clears its 1\% false positive threshold, so there is evidence for a companion, but the localization conditions above are not met -- the competitive region is wide or pushed out against the long-period end of the grid, or the significance comes from the acceleration channel alone. This is the periodogram's counterpart of a companion with only a detectable orbit.

  \item \textbf{Significant peak, period localized.} The periodogram channel itself clears its 1\% false positive threshold and both localization conditions hold, so that the period is pinned down to better than the tolerance above. Localization is never claimed on the strength of the acceleration channel. This is the periodogram's counterpart of a companion with a characterizable orbit, though a localized period is a necessary rather than a sufficient condition for a companion's orbit to be characterized. 
\end{itemize}

No injected quantity enters this classification at any stage, and the identical procedure can therefore be run on real DR4 epoch astrometry.

Figure~\ref{fig:examples} shows the three periodogram classifications for 10,000 one-companion systems in companion mass vs. period space and highlights 9 sample systems for which we also show the periodogram results. This illustrates the characteristic failure mode of the non-localized population. Beyond the observing baseline the $\Delta\chi^2$ curve does not break into competing peaks, but instead rises into a plateau pushed out against the long-period end of the periodogram grid -- providing only a lower limit on the period. Figure~\ref{fig:characterizability} shows the classification across all four companion populations.

The solid curves in Figure~\ref{fig:examples} and Figure~\ref{fig:characterizability} are again loci of constant ${\rm SNR}_{\rm tot}$. We evaluate the ${\rm SNR}_{\rm tot}$ locus star by star and plot the median, which is the mass at which half the sample reaches a given signal, rather than substituting median parameters into the formula. This ${\rm SNR}_{\rm tot}$ is the same proxy used to define the high-SNR populations (Eq.~\ref{eq:snrtot}) and it is encouraging that the recovered companions trace a similar shape as the ${\rm SNR}_{\rm tot}$ contours.

We can see in Figure~\ref{fig:characterizability} that period localization begins to truncate at the \textit{Gaia} DR4 baseline, while signal recovery does not. Inside the baseline a companion that clears the significance threshold often also has its period pinned down, but past $T_{\rm baseline}$ the two come apart (see Section~\ref{sec:results_inj_rec}). Once the period exceeds $T_{\rm baseline}$, what survives the five-parameter fit is the curvature of the observed arc. Increasing the companion mass thus moves a system across the significance boundary but not across the localization one. 

\subsection{\citet{LammersWinn_GaiaForecast} DR4 forecast}
\label{sec:results_lw25}

\citet{LammersWinn_GaiaForecast} (henceforth LW26) forecast \textit{Gaia} DR4 planet yields by combining simulated epoch astrometry via \texttt{gaiamock} \citep{El-Badry2024} with an assumed occurrence function from the California Legacy Survey \citep[CLS;][]{Rosenthal2021, Fulton2021}. To construct their mock catalog, they fit each simulated source first via an optimization step followed by MCMC analysis for systems with strong evidence for a companion. They deem a planet detected when three criteria are met together: the one-companion model is favored by $\Delta\chi^2 > 50$; the best-fit companion mass is below $13\,M_{\rm Jup}$; and the orbital period is constrained to at least $50\%$ precision (i.e., $P_{84}/P_{16} < 1.5$), where $P_{16}$ and $P_{84}$ are the 16th and 84th percentiles of the period posterior, so that their ratio spans the central 68\% credible interval. In total, $7{,}545$ of their simulated planets satisfy all three criteria. They also highlight a subset of $1{,}938$ planets whose period and mass are each recovered to $20\%$ ($P_{84}/P_{16}$ and $M_{84}/M_{16}$ both $< 1.2$). These $1{,}938$ are the planets LW26 forecast to have \emph{characterized orbits}.

The authors also made available the full set of fits rather than only the planets that survive those criteria, with posterior percentiles for the period, mass, eccentricity and inclination of every system.\footnote{\url{https://github.com/CalebLammers/GaiaForecasts}} That file contains $29{,}129$ systems, every one of which satisfies their first criterion ($\Delta\chi^2 > 50$), so it is precisely the population for which their pipeline finds significant evidence of a companion. Setting aside the $1{,}938$ systems whose period and mass are each measured to $20\%$ leaves $27{,}191$ systems whose signals their analysis identifies as significant but whose orbits it does not measure to that precision; these are the planets LW26 forecast to have only \emph{detected orbits}. Of these $27{,}191$ planets with a detected signal but whose orbit is not characterized, $18{,}778$ of the planets do not have their period measured to $20\%$; a further $8{,}244$ do, but not their mass; and the remaining $169$ have both measured to $20\%$ but a best-fit mass above $13\,M_{\rm Jup}$. Since every companion in their simulation was injected below that mass by construction, this last group are not brown dwarfs but planets whose fitted mass overshot the cut. We note, additionally, that LW26 restricted their injected signals to periods shorter than $7\,$yr because their forecasts were focused on well-characterized planets, so companions in the regime where localization fails most severely are absent from the forecast altogether, and thus this count is quite conservative.

\subsection{Injected vs. Recovered Companion Periods}
\label{sec:results_inj_rec}

\begin{figure*}
\centering
\includegraphics[width=\textwidth]{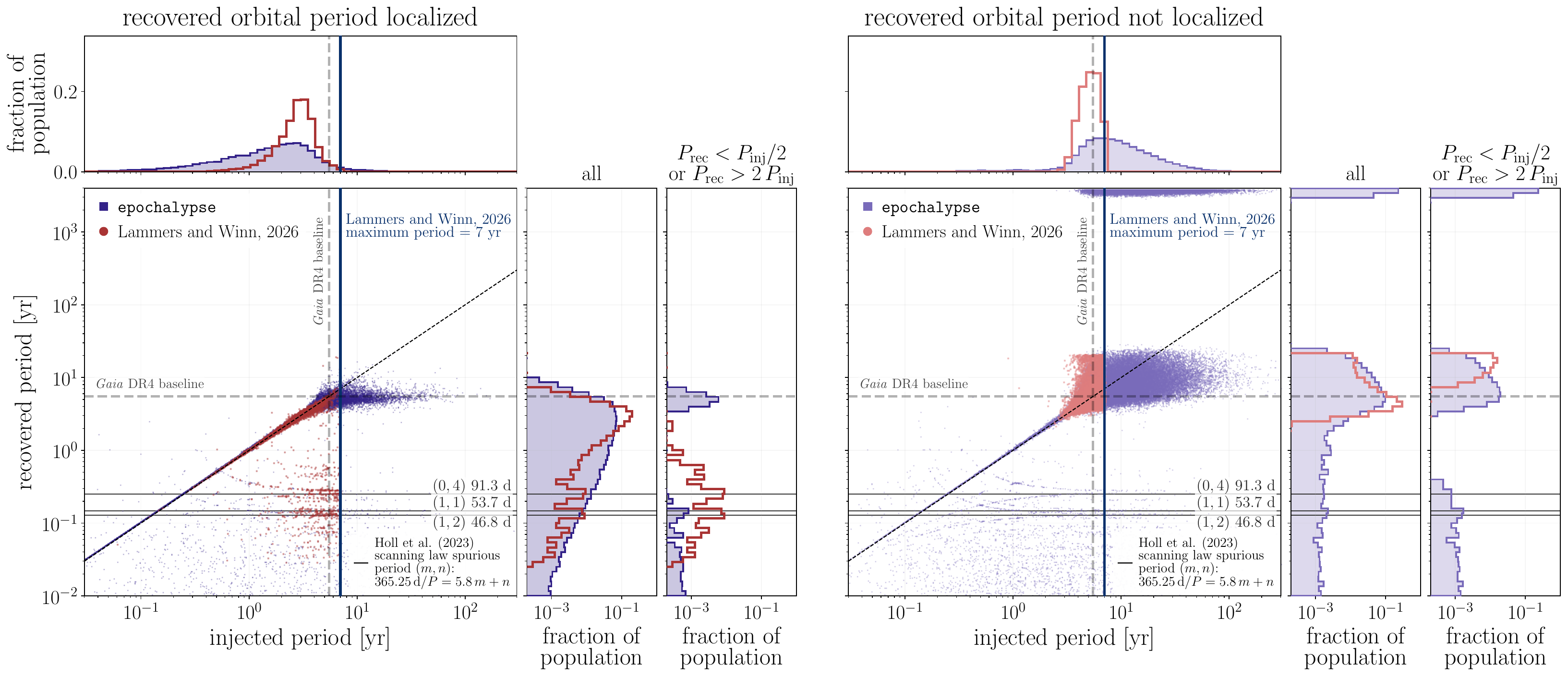}
\caption{Injected versus recovered period for the \texttt{epochalypse} one-companion: high-SNR population and the \citet{LammersWinn_GaiaForecast} \textit{Gaia} DR4 forecast, split by whether the recovered period is localized (left) or not (right). For \texttt{epochalypse}, only companions with a significant periodogram peak are shown, classified as in Section~\ref{sec:results_periodogram}. For \citet{LammersWinn_GaiaForecast}, the period is taken as localized when their posterior satisfies $P_{84}/P_{16} < 1.2$. All histograms are normalized to the sample total. The dashed diagonal marks $P_{\rm rec} = P_{\rm inj}$, and the dashed grey lines mark the \textit{Gaia} DR4 baseline on both axes. The solid navy line marks 7~yr, beyond which \citet{LammersWinn_GaiaForecast} injected no planets. Solid black lines mark three of the scanning-law spurious periods predicted by \citet{Holl2023_scanlaw}, $365.25\,{\rm d}/P = 5.8\,m + n$.}
\label{fig:injected_recovered_periods}
\end{figure*}

As a first test of the \texttt{epochalypse} catalog -- and of the \texttt{kepmodel} periodogram analysis of Section~\ref{sec:results_periodogram} -- we compare the injected and recovered periods of every one-companion: high-SNR system with a significant periodogram peak, and the forecasted planets of LW26. Figure~\ref{fig:injected_recovered_periods} separates both samples on period localization. 

For \texttt{epochalypse} we use the classification as described in Section~\ref{sec:results_periodogram}, so in Figure~\ref{fig:injected_recovered_periods} the left panel contains the ``significant peak, period localized'' class and the right panel the significant peaks of the ``period not localized'' class; the $42{,}771$ companions significant through the acceleration channel alone have no recovered period and are not shown. In the \texttt{epochalypse} high-SNR one-companion population, $204{,}308$ significant peaks divide into $106{,}339$ localized and $97{,}969$ not localized. 

For LW26 we split on their period criterion, $P_{84}/P_{16}\,<\,1.2$: the left panel holds the $10{,}351$ planets that meet it (the $1{,}938$ forecasted characterized planets together with the $8{,}244 + 169$ whose period is measured but whose mass fails on their two mass cuts), and the right panel the $18{,}778$ that do not. 

In what follows a recovered period is \emph{correct} when it is within $20\%$ of the injected period. We again emphasize that the \texttt{epochalypse} catalog does not include a realistic companion occurrence rate model but rather companions spanning wide-ranging masses and orbits, so these numbers are relevant only as a comparison between each other and should not be taken as a quantitative forecast of DR4 yield. We highlight several features of Figure~\ref{fig:injected_recovered_periods}:

\begin{enumerate}
    \item \textbf{Range of injected periods.} The injected periods of recovered \texttt{epochalypse} systems, whether or not that period is localized, extend to much shorter and much longer periods than those of LW26: the central $99\%$ of the \texttt{epochalypse} sample spans $0.045$--$68$~yr, versus $0.7$--$7.0$~yr for LW26, and $30\%$ of the \texttt{epochalypse} systems have injected periods beyond LW26's 7~yr maximum. Two choices in the injected populations drive this. First, \texttt{epochalypse} extends to $80\,M_{\rm Jup}$, while LW26 limit their injections to planetary masses and thus a $13\,M_{\rm Jup}$ maximum. Second, LW26 adopt a maximum injected period of 7~yr, whereas \texttt{epochalypse} extends companions out to $100$~au. In both the short- and long-period regimes the more massive substellar companions dominate the detection space.

    \item \textbf{Localized periods (left panel).} Companions with localized periods are preferentially those with injected periods shorter than the \textit{Gaia} DR4 baseline of $5.5$~yr, and their recovered periods are overwhelmingly correct. Of the localized \texttt{epochalypse} systems, $94\%$ have injected periods below $T_{\rm baseline}$ ($98\%$ for LW26), and $93\%$ have a correct recovered period ($92\%$ for LW26). 

    \item \textbf{Non-localized periods (right panel).} Companions whose signal is significant but whose period is not localized are preferentially those with injected periods longer than $T_{\rm baseline}$: $73\%$ of the non-localized \texttt{epochalypse} systems lie beyond it ($39\%$ for LW26, whose injections stop at 7~yr). Beyond the baseline the recovered period is usually wrong. For injected periods above $T_{\rm baseline}$, $83\%$ of \texttt{epochalypse} significant peaks recover a period off by more than $20\%$ as do $67\%$ of the LW26 forecasted planets. Overall only $32\%$ of the non-localized \texttt{epochalypse} periods are correct ($57\%$ for LW26).

    \item \textbf{The periodogram grid edge.} A substantial subset of the non-localized \texttt{epochalypse} systems, $27{,}188$ or $28\%$ of that class, is recovered at the long-period end of the \texttt{kepmodel} period grid, $P_{\rm rec} \simeq 3{,}100$--$3{,}700$~yr, the plateau seen in the periodograms in Figure~\ref{fig:examples} pushed out against the grid boundary. Their injected periods lie essentially entirely ($27{,}178$ of $27{,}188$) beyond the $T_{\rm baseline}/\sqrt{2} \simeq 3.9$~yr predicted maximally recoverable astrometric period and $97\%$ beyond $T_{\rm baseline}$ itself. 

    \item \textbf{Scanning-law spurious periods.} Finally, a subset of systems in both catalogs is recovered at errantly short periods, $P_{\rm rec} < P_{\rm inj}/2$: $6.1\%$ of the localized LW26 planets ($628$) and $2.9\%$ of the localized \texttt{epochalypse} systems ($3{,}098$). These spurious signals are often recovered at the spurious periods that \citet{Holl2023_scanlaw} predicted to be induced by the \textit{Gaia} scanning law, at periods $P$ given by
    \begin{equation}
        365.25\,{\rm d}/P = 5.8\,m + n,
    \end{equation}
    \noindent
    where $m$ and $n$ are small integers and 5.8 cycles yr$^{-1}$ (approximately 63~d) is the precession frequency of the \textit{Gaia} spin axis around the Sun during the nominal scanning law. $40\%$ of the errant localized LW26 periods lie within $3\%$ of one of these lines, with the strongest excesses at $(m,n) = (1,2)$, $46.8$~d, and $(0,4)$, $91.3$~d. The same lines are present in the errant localized \texttt{epochalypse} systems ($9\%$ of the errant localized periods in total), with significant excesses at $(m,n) = (1,2)$, $46.8$~d and at $(1,1)$, $53.7$~d. It is worth noting that in the $1{,}938$ orbit characterized forecasted systems in LW26, there are no errant scanning law induced spurious signals as the two mass cuts move all spurious recovered signals into the orbit detected-not-characterized regime.
\end{enumerate}

\subsection{Substellar Companions in \textit{Gaia} DR4} \label{sec:compare_companions}
\begin{figure*}
\centering
\includegraphics[width=\textwidth]{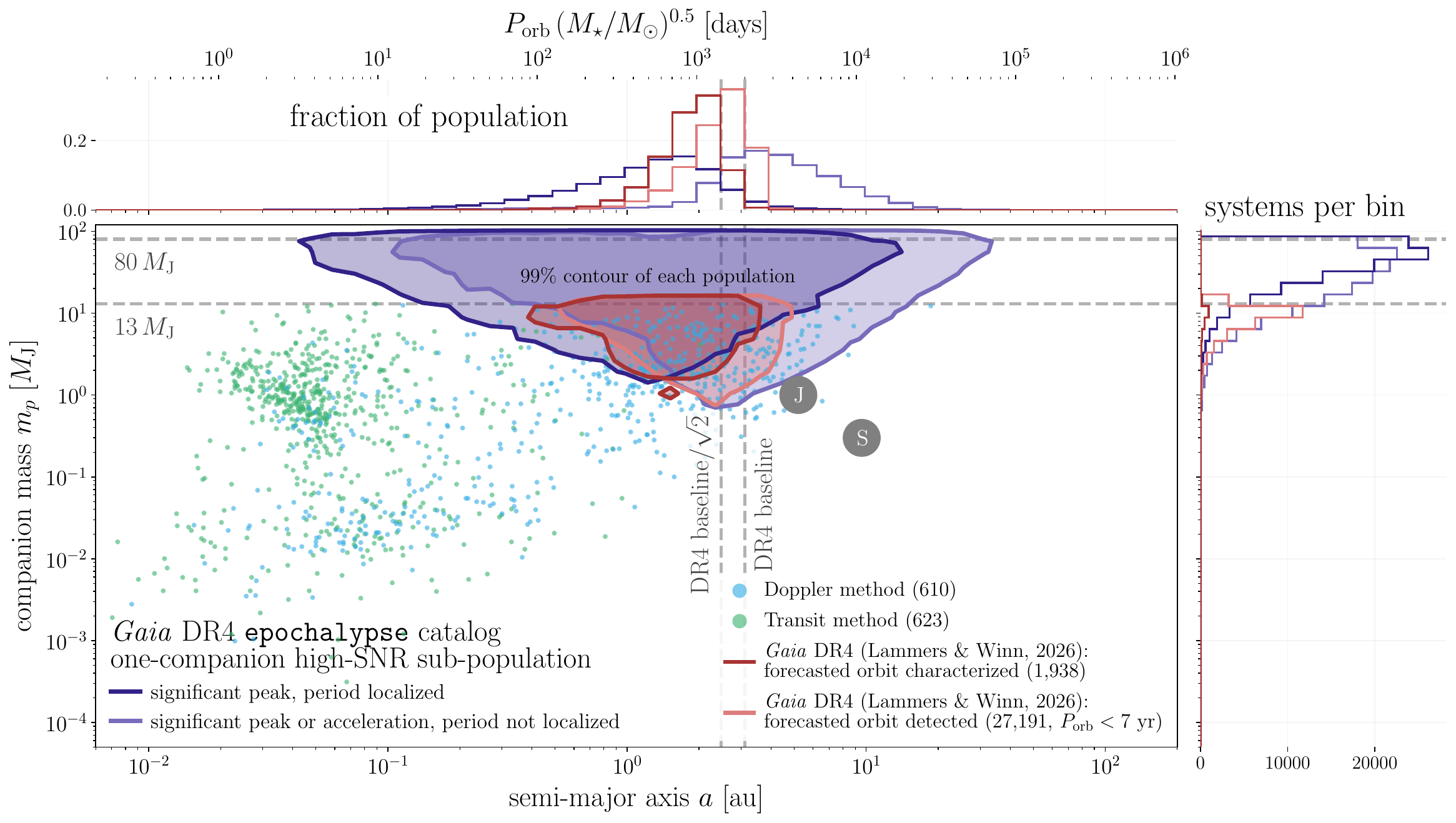}
\caption{Companion mass versus semi-major axis for the \texttt{epochalypse} one-companion: high-SNR population, the \citet{LammersWinn_GaiaForecast} \textit{Gaia} DR4 forecast, and confirmed Doppler and transit planets. The \texttt{epochalypse} companions are split via their periodogram classification (localized vs. not localized periods), and the \citet{LammersWinn_GaiaForecast} planets by whether their MCMC posterior characterizes the orbit ($P_{84}/P_{16} < 1.2$, $M_{84}/M_{16} < 1.2$, and best-fit mass $< 13\,M_{\rm J}$). \citet{LammersWinn_GaiaForecast} injected planets only out to 7~yrs so $27{,}191$ is a lower limit, not a prediction of number of orbit detected but not characterized planets. \textit{Gaia} forecasts shown are 99\% contours and histograms. Dashed horizontal lines mark the $13\,M_{\rm J}$ mass cut applied in \citet{LammersWinn_GaiaForecast} and the $80\,M_{\rm J}$ ceiling of the \texttt{epochalypse} injected population. The dashed verticals mark the \textit{Gaia} DR4 baseline and the baseline divided by $\sqrt{2}$, which is the optimal period for companion recovery predicted by ${\rm SNR}_{\rm tot}$ (Eqs.~\ref{eq:snrtot} and \ref{eq:popt}) and intersects near the minimum of both detected but not localized contours.}
\label{fig:lw25_comparison}
\end{figure*}

Figure~\ref{fig:lw25_comparison} places the \texttt{epochalypse} one-companion: high-SNR population in the same plane as the \citet{LammersWinn_GaiaForecast} forecasted characterized and detected planets. We show the 99\% 2D contours and histograms in companion period and companion mass space. Since our populations carry no occurrence model, the density of \texttt{epochalypse} companions reflects our log-uniform injection distribution, and ratios between our counts and LW26 would not be meaningful. What is comparable is where each class falls. The companions whose periods our periodogram localizes overlap with their planets with characterized orbits; while those that are significant but not localized extend to larger separations along the same lower envelope, in the region their planets with only detected orbits occupy. It is worth emphasizing that in the LW26 forecasts, there are $\sim$14 times more planets with their orbits detected than with their orbits characterized ($27{,}191$ vs. $1{,}938$) and this number is a lower limit as LW26 truncates injection at 7~yrs. Thus, this population of companions that are detected but uncharacterized will be significant in \textit{Gaia} DR4 and will extend to both smaller masses and longer orbital periods than the characterizable population.

\section{Using the \texttt{epochalypse} catalog} \label{sec:use}

The catalog is archived on Zenodo \citep{yahalomi2026_epochalypse_zenodo} at \dataset[doi:10.5281/zenodo.23789041]{https://doi.org/10.5281/zenodo.23789041}, and the generation pipeline, the periodogram characterization code, the figure notebooks and the tutorial are at \url{https://github.com/dyahalomi/epochalypse}. The catalog is designed to support orbit-modeling, detection-completeness studies, false-positive validation, and end-to-end benchmarking of \textit{Gaia} astrometric planet-recovery pipelines. Once \textit{Gaia} DR4 is released and we have a more complete understanding of the noise model and systematics of the epoch astrometry in the DR4, we envision that users may want to inject signals (i.e., substellar companions) onto the real noise. Here we have presented not only a mock catalog based on our current best predictions of the DR4 noise, but also a framework (and the underlying code) for injecting substellar signals onto DR4 noise, once it is publicly available.

\subsection{Data products and access} \label{sec:use_data}

\begin{deluxetable*}{llll}
\tablecaption{\texttt{epochalypse} simulated DR4 epoch astrometry (one row per field-of-view (FoV) transit; files \texttt{simulated\_astrometry/<population>/epochs\_rankNNNNN\_of\_00320.parquet}).\label{tab:astrometry}}
\tablewidth{0pt}
\tabletypesize{\scriptsize}
\tablehead{
  \colhead{Column} & \colhead{Unit} & \colhead{Description} & \colhead{References}
}
\startdata
\cutinhead{Epoch sampling (from the \textit{Gaia} scanning law)}
\texttt{obs\_time\_tcb}       & d (JD, TCB) & Observation epoch $t_{\rm TCB}$ of the FoV transit & Sec.~\ref{sec:scanlaw1} \\
\texttt{scan\_pos\_angle}     & rad         & Scan angle $\psi$, in $(-\pi, \pi]$ & Eq.~\ref{eq:5param} \\
\texttt{parallax\_factor\_al} & $\cdots$    & Along-scan parallax factor $\Pi$ & Eq.~\ref{eq:5param} \\
\texttt{field\_of\_view}      & $\cdots$    & \texttt{FoVP} (preceding) or \texttt{FoVF} (following) & Sec.~\ref{sec:scanlaw1} \\
\cutinhead{Simulated measurement}
\texttt{centroid\_pos\_al}        & mas & Along-scan position $w_{\rm obs} = w_{\rm 5par} + w_{\rm reflex} + \varepsilon$ & Eqs.~\ref{eq:5param}, \ref{eq:reflex}, \ref{eq:sigma_injected} \\
\texttt{centroid\_pos\_error\_al} & mas & Reported $1\sigma$ uncertainty $\sigma_{{\rm reported},i}$ & Eq.~\ref{eq:sigma_reported} \\
\cutinhead{Identifiers and bookkeeping}
\texttt{system\_id}       & $\cdots$ & \texttt{<population>\_<gaia\_source\_id>}; joins to Table~\ref{tab:catalog} & Sec.~\ref{sec:astrometry} \\
\texttt{gaia\_source\_id} & $\cdots$ & \textit{Gaia} DR3 source identifier of the host star & \textit{Gaia} DR3 \\
\texttt{source\_id\_dr2}  & $\cdots$ & \textit{Gaia} DR2 source identifier of the host star & \textit{Gaia} DR2 \\
\texttt{system\_seed}     & $\cdots$ & Per-system astrometry seed (Table~\ref{tab:catalog}) & Sec.~\ref{sec:astrometry} \\
\enddata
\tablecomments{Each population directory (\texttt{0\_companion}, \texttt{1\_companion}, \texttt{2\_companion}) holds 320 shards; the \texttt{0\_companion} set is the companion-free control. Each row is one along-scan measurement at an epoch. The measurement model is the five-parameter astrometric solution (Eq.~\ref{eq:5param}) plus the companion-induced reflex (Eq.~\ref{eq:reflex}), projected onto the along-scan direction, with noise $\varepsilon \sim \mathcal{N}(0, \sigma_{{\rm injected},i}^2)$ (Eq.~\ref{eq:sigma_injected}). The reported uncertainty is the smaller $\sigma_{{\rm reported},i}$ (Eq.~\ref{eq:sigma_reported}); both carry the same 10\% per-epoch jitter (Sec.~\ref{sec:noise}). The three identifier columns are stored as strings; all other columns are 64-bit floats or integers.}
\end{deluxetable*}

\begin{deluxetable*}{llll}
\tablecaption{\texttt{epochalypse} injected-solutions catalog (one row per system; files \texttt{truths\_rankNNNNN\_of\_00320.parquet} in the population directories of Table~\ref{tab:astrometry}, concatenated per population into \texttt{injected\_solutions\_<population>.parquet} and \texttt{injected\_solutions\_<population>\_high\_snr.parquet}).\label{tab:catalog}}
\tablewidth{0pt}
\tabletypesize{\scriptsize}
\tablehead{
  \colhead{Column} & \colhead{Unit} & \colhead{Description} & \colhead{References}
}
\startdata
\cutinhead{Identifiers and seeds}
\texttt{system\_id}       & $\cdots$ & \texttt{<population>\_<gaia\_source\_id>}; joins to Table~\ref{tab:astrometry} & Sec.~\ref{sec:astrometry} \\
\texttt{population}       & $\cdots$ & \texttt{0\_companion}, \texttt{1\_companion}, or \texttt{2\_companion} & Sec.~\ref{sec:companions} \\
\texttt{gaia\_source\_id} & $\cdots$ & \textit{Gaia} DR3 source identifier of the host star & \textit{Gaia} DR3 \\
\texttt{source\_id\_dr2}  & $\cdots$ & \textit{Gaia} DR2 source identifier of the host star & \textit{Gaia} DR2 \\
\texttt{n\_transits\_dr4} & $\cdots$ & Number of DR4 FoV transits (rows in Table~\ref{tab:astrometry}) & Sec.~\ref{sec:scanlaw1} \\
\texttt{master\_seed\_planets}    & $\cdots$ & Global seed for the companion draws (42) & Sec.~\ref{sec:companions} \\
\texttt{master\_seed\_astrometry} & $\cdots$ & Global seed for the astrometric noise (45) & Sec.~\ref{sec:astrometry} \\
\texttt{system\_seed}      & $\cdots$ & Per-system astrometry seed\tablenotemark{b} & Sec.~\ref{sec:astrometry} \\
\texttt{noise\_seed}       & $\cdots$ & Seed for the 10\% per-epoch jitter on the uncertainties & Eqs.~\ref{eq:sigma_injected}, \ref{eq:sigma_reported} \\
\texttt{observation\_seed} & $\cdots$ & Seed for the along-scan noise $\varepsilon$ & Sec.~\ref{sec:astrometry} \\
\cutinhead{Host-star properties}
\texttt{parallax\_mas}    & mas            & Host-star parallax (\texttt{G23H}) & \textit{Gaia} DR3 \\
\texttt{pmra\_mas\_yr}    & mas\,yr$^{-1}$ & Host proper motion in R.A. (\texttt{G23H}) & \textit{Gaia} DR3 \\
\texttt{pmdec\_mas\_yr}   & mas\,yr$^{-1}$ & Host proper motion in decl. (\texttt{G23H}) & \textit{Gaia} DR3 \\
\texttt{mass\_st\_msun}   & $M_\odot$      & Stellar mass, interpolated from spectral type & Sec.~\ref{sec:pecaut} \\
\texttt{radius\_st\_rsun} & $R_\odot$      & Stellar radius, interpolated from spectral type & Sec.~\ref{sec:pecaut} \\
\texttt{sigma\_single\_mas} & mas          & Per-transit injected uncertainty $\sigma_{\rm UEVA,single}$ (no jitter) & Eq.~\ref{eq:sigma_UEVA_single} \\
\texttt{n\_planets}       & $\cdots$       & Number of injected companions (0, 1, or 2) & Sec.~\ref{sec:companions} \\
\cutinhead{Injected (truth) orbital elements\tablenotemark{a}}
\texttt{sma\_1}, \texttt{sma\_2}         & au        & Semimajor axis $a$ & Sec.~\ref{sec:companions} \\
\texttt{ecc\_1}, \texttt{ecc\_2}         & $\cdots$  & Eccentricity $e$ & Sec.~\ref{sec:companions} \\
\texttt{inc\_1}, \texttt{inc\_2}         & deg       & Inclination $i$ & Sec.~\ref{sec:companions} \\
\texttt{Omega\_1}, \texttt{Omega\_2}     & deg       & Longitude of the ascending node $\Omega$ & Sec.~\ref{sec:companions} \\
\texttt{omega\_1}, \texttt{omega\_2}     & deg       & Argument of periastron $\omega$ & Sec.~\ref{sec:companions} \\
\texttt{M\_anom\_1}, \texttt{M\_anom\_2} & deg       & Mean anomaly $M_0$ at $t_0 = \mathrm{JD}\ 2457936.875$ (TCB) & Sec.~\ref{sec:companions} \\
\texttt{mass\_pl\_1}, \texttt{mass\_pl\_2} & $M_{\rm Jup}$ & Companion mass $M_p$ & Sec.~\ref{sec:companions} \\
\texttt{period\_1}, \texttt{period\_2}   & yr        & Orbital period $P = \sqrt{a^{3}/(M_\star + M_p)}$, masses in $M_\odot$ & Sec.~\ref{sec:companions} \\
\cutinhead{Detectability metrics\tablenotemark{a}}
\texttt{alpha\_mas\_1}, \texttt{alpha\_mas\_2}   & mas      & Stellar reflex semimajor axis $\alpha$ & Eq.~\ref{eq:snr1} \\
\texttt{snr\_single\_1}, \texttt{snr\_single\_2} & $\cdots$ & Single-transit SNR, $\alpha/\sigma_{\rm UEVA,single}$ & Eq.~\ref{eq:snr1} \\
\texttt{snr\_eff\_1}, \texttt{snr\_eff\_2}       & $\cdots$ & Period-suppressed single-transit SNR & Eq.~\ref{eq:snreff} \\
\texttt{snr\_total\_1}, \texttt{snr\_total\_2}   & $\cdots$ & Total SNR over all DR4 transits & Eq.~\ref{eq:snrtot} \\
\cutinhead{Two-companions architecture (\texttt{2\_companion} only)}
\texttt{P\_ratio}        & $\cdots$ & Period ratio $P_2/P_1$ (outer to inner) & Sec.~\ref{sec:companions} \\
\texttt{near\_2\_1}      & $\cdots$ & Flag: $P_2/P_1$ within 5\% of 2:1 & Eq.~\ref{eq:mmr} \\
\texttt{near\_3\_2}      & $\cdots$ & Flag: $P_2/P_1$ within 5\% of 3:2 & Eq.~\ref{eq:mmr} \\
\texttt{near\_resonance} & $\cdots$ & Flag: \texttt{near\_2\_1} $\lor$ \texttt{near\_3\_2} & Eq.~\ref{eq:mmr} \\
\texttt{coplanar}        & $\cdots$ & Flag: outer orbit shares the inner orbit's $i$ and $\Omega$ & Sec.~\ref{sec:companions} \\
\enddata
\tablecomments{Host-star parallax and proper motions are the \textit{Gaia} DR3 values carried in the \texttt{G23H} catalog \citep{Thompson_G23H}; stellar mass and radius are interpolated from the \citet{PecautMamajek2013} table (Sec.~\ref{sec:pecaut}).}
\tablenotetext{a}{Stored per companion with the suffix \texttt{\_1} (inner) and
  \texttt{\_2} (outer), ordered by semimajor axis. Columns are present only where
  they apply: the \texttt{0\_companion} table has no companion columns, the
  \texttt{1\_companion} table has only the \texttt{\_1} columns, and the
  \texttt{2\_companion} table has both sets plus the architecture flags. The
  high-SNR tables keep the base population name in the \texttt{population} column.
  The three identifier columns are stored as strings, the flags as booleans, and
  all other columns as 64-bit floats or integers.}
\tablenotetext{b}{Derived deterministically from \texttt{master\_seed\_astrometry}, \texttt{population},
  and \texttt{gaia\_source\_id}; parent of \texttt{noise\_seed} and \texttt{observation\_seed}.
  The companion draws use a per-system seed derived in the same way from
  \texttt{master\_seed\_planets}; it is not stored but is reproducible.}
\end{deluxetable*}

For every system in \texttt{epochalypse} we store a \textit{Gaia} DR4-like epoch table with one row per epoch, containing the observation epoch, the along-scan centroid position and its reported formal uncertainty, the parallax factor, the scan angle, the field-of-view identifier, and the \textit{Gaia} DR3/DR2 source identifiers (see Section~\ref{sec:astrometry} and Table~\ref{tab:astrometry}).

For each simulated system we release the epoch-by-epoch astrometry along with the
true companion parameters used to generate it. These are stored under
\texttt{simulated\_astrometry/}, in one directory per population
(\texttt{0\_companion}, \texttt{1\_companion}, \texttt{2\_companion}). Each directory
holds 640 compressed Parquet files: the astrometry in
\texttt{epochs\_rankNNNNN\_of\_00320.parquet} and the matching companion parameters in
\texttt{truths\_rankNNNNN\_of\_00320.parquet}, where \texttt{NNNNN} runs from
\texttt{00000} to \texttt{00319}. The split into 320 pieces reflects the number of
processes (shards) used to generate the simulations and carries no physical meaning;
corresponding \texttt{epochs} and \texttt{truths} files share a rank and can be joined
on the system identifier:
\begin{itemize}
  \item \texttt{epochs\_rank<NNNNN>\_of\_00320.parquet} -- the epoch tables (see Table~\ref{tab:astrometry}), one row per along-scan measurement (32 to 293 epochs across the catalog, with a median of 81). The two files of a given rank \texttt{<NNNNN>} describe exactly the same set of systems, and every row of both carries \texttt{system\_id}, \texttt{gaia\_source\_id}, and \texttt{source\_id\_dr2}, so a system's epochs are matched to its truth values by joining on \texttt{gaia\_source\_id}.
  
  \item \texttt{truths\_rank<NNNNN>\_of\_00320.parquet} -- one row per system, listing the injected stellar parameters (parallax, proper motions, mass, radius) and each companion's orbital elements ($a$, $e$, $i$, $\Omega$, $\omega$, $M_0$, $M_p$), together with the derived detectability metrics and the random seeds used to generate it (see Table~\ref{tab:catalog});
\end{itemize}

For convenience the 320 truth shards of each population are also concatenated and sorted by \texttt{gaia\_source\_id} into a single table\footnote{\texttt{injected\_solutions\_<population>.parquet}} ($5{,}040{,}328$ systems each), alongside the high-SNR sub-populations,\footnote{\texttt{injected\_solutions\_<population>\_high\_snr.parquet}} which retain the systems whose every companion has $\mathrm{SNR}_{\rm tot} \geq 5$ ($439{,}503$ and $26{,}379$ systems for the one- and two-companions populations, respectively). The stellar sample and the scanning-law predictions from which the epochs were built are released with them, as \texttt{stars.csv} and \texttt{index/stars\_index.parquet}, \texttt{index/scanlaw\_index.parquet}.

Because every system carries the seeds and injected parameters that produced it, any solution can be regenerated exactly, and every fit can be scored directly against truth. The layout uses only standard Parquet files with flat, primitive column types, so both tables can be read -- whole, by column, or filtered on \texttt{gaia\_source\_id} -- with any Parquet implementation.

\subsection{Orbit-modeling tutorial}

The repository's tutorial modeling notebook, \texttt{fit\_astrometric\_orbits.ipynb}, is the recommended entry point. It loads a single simulated system from the \texttt{epochalypse} catalog, reconstructs the along-scan design matrix (the five-parameter astrometric model of Section~\ref{sec:astrometry}) together with a \texttt{jaxoplanet} Keplerian reflex model, and fits the same epoch data with zero-, one-, and two-planet models using Hamiltonian Monte Carlo via the ``No-U-Turn Sampler'' (NUTS) in \texttt{NumPyro} \citep{numpyro2, numpyro1}. The competing models are compared with the widely applicable information criterion \citep[WAIC;][]{Watanabe2010}, and the recovered period and mass posteriors are plotted against the injected truth.

The configuration cell contains two pre-identified, high-signal example systems -- one single-companion and one two-companions -- re-derived from the injected-solutions table so that they remain valid across data regenerations. Selecting a different population and system index reproduces the same workflow for any system in the catalog.

\subsection{Potential use cases}

Here we outline a few of the ways in which we envision the \texttt{epochalypse} catalog being used (though there are likely others).

The catalog is built to test and compare \textit{Gaia} DR4 modeling pipelines. The epoch tables are written in the same format as the expected \textit{Gaia} DR4 epoch astrometry products, so any package built to model \textit{Gaia} DR4 data can be run natively on \texttt{epochalypse} and its posteriors compared to the injected values. We expect sampler comparison to matter most for the companions with only ``detectable'' orbits of Section~\ref{sec:results}, i.e., systems where a companion is clearly present in the data but the period and thus orbit is not pinned down. This is where the choices of prior, parameterization, and sampler are likely to have the largest effect on the solution.

The catalog also enables analyses of where orbit recovery breaks down. For each system in the one-companion population one can compare the recovered period, mass, and eccentricity to the injected values, and then ask how the agreement depends on, e.g., $\alpha$, $P/T_{\rm baseline}$, $e$, or the number of transits. We injected companions across the full parameter space with no detectability cut, so this comparison includes systems for which no orbit can be recovered.

We did not build any occurrence model into the catalog, so the recovery rates describe the measurement and not intrinsic occurrence rate. However, a user can reweight them to whatever occurrence model they prefer (e.g., via importance sampling) in order to forecast a DR4 yield or to correct one. The high-SNR sub-populations are useful here as they are dense in systems near and above the detection threshold.

The two-companions population allows for tests of joint multi-Keplerian fits, including the recovery of period ratios and mutual inclinations (coplanar and mutually inclined systems are flagged in the catalog). We retained near-resonant systems (Section~\ref{sec:companions}) in part so that users can test whether their sampler is able to tell a single eccentric companion apart from a $2{:}1$ resonant pair \citep{Yahalomi_resoeccentric, Greenfield2026}.

Lastly, we note again that we release the full generation pipeline along with the catalog. Once DR4 epoch astrometry is public, the same companion populations can be injected on top of real DR4 noise, and all of the tests above can be repeated with the actual systematics in place.

\section{Conclusion} \label{sec:conclusions}

\textit{Gaia} DR4 will deliver epoch astrometry for billions of sources, and the orbit-modeling tools built for it have yet to be tested against a common set of systems with known answers. We present the \texttt{epochalypse} catalog: \textit{Gaia} DR4-like epoch astrometry for zero-, one-, and two-companions systems injected onto every one of the $5{,}040{,}328$ main-sequence stars within $500\,{\rm pc}$ in the \texttt{G23H} catalog, with companions drawn log-uniformly over $1\,M_{\rm Mars}$--$80\,M_{\rm Jup}$ and $0.03$--$100$~au, no detectability cut, and no occurrence model (Sections~\ref{sec:stars}--\ref{sec:astrometry}). The zero-companions population is a matched null for the other two, and high-SNR sub-populations flag the $439{,}503$ one-companion and $26{,}379$ two-companions systems with ${\rm SNR}_{\rm tot} \geq 5$. Our conclusions are summarized below.

\begin{itemize}

    \item We define an effective detectability metric, ${\rm SNR}_{\rm eff} = {\rm SNR}_1 / [1 + (P/T_{\rm baseline})^2]$ (Equation~\ref{eq:snreff}), that suppresses long-period signals in proportion to the curvature of the observed arc, so that for $P \gg T_{\rm baseline}$ it reduces to the stellar acceleration times the square of the time over which it acts $\propto (M_p\,\varpi/a^2)\,T_{\rm baseline}^2$ and, with the $\sqrt{N}$ growth in transits, recovers the $T_{\rm baseline}^{5/2}$ scaling of \citet{Brandt2024}. Contours of constant ${\rm SNR}_{\rm tot}$ are V-shaped in the mass--period plane as expected, falling as $M_p \propto P^{-2/3}$ inside the baseline and rising as $M_p \propto P^{4/3}$ beyond it, with a minimum at $P_{\rm opt} = T_{\rm baseline}/\sqrt{2} \simeq 3.9$~yr. This $P_{\rm opt}/T_{\rm baseline} \simeq 0.71$ matches the $\sim$0.7 ratio at which \citet{Gould2008} found astrometric sensitivity to begin falling off via Fisher matrix analysis. The ${\rm SNR}_{\rm tot}$ predictions qualitatively match the shape of recovered signals in companion mass--period space of both \citet{LammersWinn_GaiaForecast} and periodogram recovery of \texttt{epochalypse}.

    \item Running the \texttt{kepmodel} periodogram on every system, with thresholds set on the companion-free control at a 1\% false-positive rate, we find that signal detection and period localization separate beyond the baseline. Detection follows the ${\rm SNR}_{\rm tot}$ loci of Equation~\ref{eq:snrtot}; localization fails for longer orbital periods beyond the observing baseline (Section~\ref{sec:results_periodogram}).

    \item Among high-SNR one-companion systems with localized periods, $94\%$ lie inside the baseline and $93\%$ are recovered to within $20\%$. Among those with significant but non-localized signals, $73\%$ lie beyond the DR4 baseline, and for injected periods above $T_{\rm baseline}$ the recovered period is wrong by more than $20\%$ in $83\%$ of cases. Additionally, $28\%$ are pushed to the maximum-period edge of the periodogram grid.

    \item Spurious short recovered periods, in both \texttt{epochalypse} via periodogram analysis with \texttt{kepmodel} and the \citet{LammersWinn_GaiaForecast} forecast via MCMC analysis, fall on the scanning-law induced spurious periods $365.25\,{\rm d}/P = 5.8\,m + n$ of \citet{Holl2023_scanlaw}, most strongly at $46.8$, $91.3$, and $53.7$~d. Future work should investigate whether sampling choices can help resolve this issue.

    \item In the mass--period plane, our localized companions occupy the region of the \citet{LammersWinn_GaiaForecast} characterized planets, and our non-localized companions extend along the same detection floor to larger separations. In \citet{LammersWinn_GaiaForecast} planets with detected-but-uncharacterized orbits outnumber their characterized planets by roughly fourteen to one (Section~\ref{sec:compare_companions}). DR4 will therefore contain many sources whose astrometry is clearly not that of a single star but whose period, and hence companion mass, is unmeasured. Future work should investigate whether sampling choices affect the recovery of these distant companion signals.

    \item Because every system carries its injected parameters and seeds, any DR4 pipeline can be run natively and tested on the \texttt{epochalypse} catalog. The catalog can be reweighted to an occurrence rate assumption by importance sampling to forecast or correct a yield. The two-companions population, including $16{,}634$ near-resonant pairs, can be used to test multi-Keplerian fits and the resoeccentric degeneracy \citep{Yahalomi_resoeccentric, Greenfield2026}. Once \textit{Gaia} DR4 is public, the pipeline can be used to inject the same populations onto the real noise.
\end{itemize}

The \texttt{epochalypse} catalog, generation pipeline, and orbit-modeling tutorial are public, in the hope that they prove useful as the community prepares for the \textit{Gaia} DR4 astrometric ``epochalypse.''

\software{
\texttt{astropy} \citep{astropy:2013, astropy:2018, astropy:2022},
\texttt{corner} \citep{corner},
\texttt{healpy} \citep{Gorski2005_HEALPix, Zonca2019_healpy},
\texttt{IPython} \citep{PerezGranger2007_ipython},
\texttt{JAX} \citep{jax2018github},
\texttt{jaxoplanet} \citep{jaxoplanet},
\texttt{Jupyter} \citep{Kluyver2016_jupyter},
\texttt{kepmodel} \citep{Delisle2022_kepmodel},
\texttt{matplotlib} \citep{Hunter2007_matplotlib},
\texttt{mpi4py} \citep{DalcinFang2021_mpi4py},
\texttt{NumPy} \citep{Harris2020_numpy},
\texttt{NumPyro} \citep{numpyro2, numpyro1},
\texttt{pandas} \citep{McKinney2010_pandas},
\texttt{pyarrow} \citep{pyarrow},
\texttt{Python} \citep{VanRossumDrake2009_python},
\texttt{SciPy} \citep{Virtanen2020_scipy},
\texttt{S+LEAF} \citep{Delisle2020_spleaf}
}

\begin{acknowledgements}
We are grateful to Ben Cassese, Tiger Lu, and Zephyr Penoyre for enlightening conversations and comments on the manuscript. This work started at a workshop at the Flatiron Institute, Center for Computational Astrophysics: ``Tooling up for \textit{Gaia} DR4'' and benefited from insightful conversations with the Astronomical Data Group at the Flatiron Institute, Center for Computational Astrophysics. 

This work has made use of data from the European Space Agency (ESA) mission \textit{Gaia} (\url{https://www.cosmos.esa.int/gaia}), processed by the \textit{Gaia} Data Processing and Analysis Consortium (DPAC, \url{https://www.cosmos.esa.int/web/gaia/dpac/consortium}). Funding for the DPAC has been provided by national institutions, in particular the institutions participating in the \textit{Gaia} Multilateral Agreement. In particular, this work made use of the \textit{Gaia} DR4 prerelease astrometric time series (\url{https://www.cosmos.esa.int/web/gaia/dr4-prerelease}) and of the \textit{Gaia} Observation Forecast Tool (GOST; \url{https://gaia.esac.esa.int/gost/}) to obtain the DR4 scanning law. We thank the \textit{Gaia} DPAC for making these resources available ahead of DR4.

D.A.Y. is supported by a Juan Carlos Torres Postdoctoral Fellowship at the Massachusetts Institute of Technology. C.S. acknowledges support from the Department of Energy Computational Science Graduate Fellowship under Award Number DE-SC0026073. C.D.O. acknowledges support from the B. Thomas Soifer Fellowship at the California Institute of Technology. C.L. acknowledges support from a Natural Sciences and Engineering Research Council of Canada (NSERC) Postgraduate Scholarship.

This research was enabled in part by support provided by the Digital Research Alliance of Canada (alliancecan.ca). This research used the facilities of the Canadian Astronomy Data Centre operated by the National Research Council of Canada with the support of the Canadian Space Agency.

During the preparation of this work, the authors used OpenAI's ChatGPT and Codex and Anthropic's Claude models to assist with code development and writing improvements. All AI-assisted output was reviewed, edited, and verified by the authors, who take full responsibility for the content of this manuscript.

\end{acknowledgements}


\bibliography{main}{}
\bibliographystyle{aasjournalv7}



\end{document}